\documentclass[english,12pt,sort&compress]{iopart}
\usepackage[T1]{fontenc}
\usepackage[latin1]{inputenc}
\usepackage{float}
\usepackage{iopams}
\usepackage{setstack}
\usepackage{graphicx}
\usepackage{amssymb}
\usepackage[numbers]{natbib}

\makeatletter



\usepackage{babel}
\makeatother
\begin{document}

\title[]{Phase diagram of the ABC model with nonconserving processes}

\author{A Lederhendler, O Cohen and D Mukamel}

\address{Department of Physics of Complex Systems, Weizmann Institute of Science, Rehovot 76100, Israel}

\begin{abstract}
The three species ABC model of driven particles on a ring is
generalized to include vacancies and particle-nonconserving processes.
 The model exhibits phase separation at high densities.
For equal average densities of the three species, it is shown that
although the dynamics is {\it local}, it obeys detailed balance with
respect to a Hamiltonian with {\it long-range interactions},
yielding a nonadditive free energy. The phase diagrams of the
conserving and nonconserving models, corresponding to the canonical
and grand-canonical ensembles, respectively, are calculated in the
thermodynamic limit. Both models exhibit a transition from a homogeneous
 to a phase-separated state, although the phase diagrams are shown to
differ from each other.
This conforms with the expected inequivalence of ensembles in equilibrium
systems with long-range interactions. These results are based on a stability
analysis of the homogeneous phase and exact solution of the
hydrodynamic equations of the models. They are supported by
Monte-Carlo simulations. This study may serve as a useful starting
point for analyzing the phase diagram for unequal densities, where
detailed balance is not satisfied and thus a Hamiltonian cannot be
defined.

\end{abstract}

\noindent Draft date : \today \\


\maketitle

\section{Introduction}

Systems driven out of equilibrium by an external field, such as a
temperature or a pressure gradient or by an electric field, have been
a subject of extensive studies in recent years (see \cite{jstat2007,Schmittman1995} for reviews of this topic).
 Often, such systems reach a current carrying steady
state where detailed balance is not satisfied. The nature of the
steady state typically depends on the details of the microscopic
dynamics. This is in contrast to systems in thermal equilibrium
where the equilibrium distribution is determined by the Hamiltonian,
and is independent of the microscopic dynamics. Thus, for instance,
in the Ising model, the same equilibrium state is reached in the
thermodynamic limit irrespective of whether the dynamics is magnetization conserving
(Kawasaki dynamics) or magnetization nonconserving (Glauber
dynamics).

Insight into the properties of nonequilibrium steady states has
been obtained by detailed studies of simple models of particles on a
lattice, where the particles are driven by either boundary terms or
in the bulk. The behavior of these driven systems has been shown to
be nontrivial, as they exhibit phenomena which do not occur in
equilibrium. In particular, nonequilibrium steady states of driven
systems have been found to exhibit long-range correlations,
even for systems with strictly local dynamics
\cite{Spohn1983,Dorfman1994,OrtizDeZarate2004,Bertini2007b,Derrida2007,Bodineau2008}. While attempting to
understand the nature of these long-range correlations, several
studies revealed similarities
between some properties of driven nonequilibrium systems and
equilibrium systems with long-range interactions. One example is the
presence of phase transitions in one-dimensional driven
models \cite{Evans2000,Mukamel2000,Kafri2002}, which are known to exist also
in equilibrium systems with long-range interactions.
An interesting question, which is addressed in the present
paper, is to what extent can studies of equilibrium long-range
systems provide some understanding of properties of driven
nonequilibrium systems.

Equilibrium systems with long-range interactions are those where
 the two body potential decays at large distance, $r$, as $1/r^{d+\sigma}$, with
$-d\leq\sigma\leq0$ in $d$ dimensions. In this case the total energy
of a system of homogeneously distributed particles scales as
$V^{1-\sigma/d}$ with its volume $V$. Therefore, if
$-d\leq\sigma\leq0$, the energy scales super-linearly with $V$, and
the system is nonadditive. As a result of the nonadditivity the
various statistical mechanical ensembles may become inequivalent
\cite{Antonov1962,LyndenBell1968,Thirring1970,Hertel1971,LyndenBell1999,
Thirring2003,Barre2001,Barre2002,Mukamel2005}.
For example, it is well known that the specific heat of long-range
interacting systems can be negative in the microcanonical ensemble.
On the other hand within the canonical ensemble the specific heat
is proportional to the energy fluctuations,
$C_V\propto\langle E^{2}\rangle-\langle E\rangle^{2}$, and always
positive. Similar effects can be found when comparing the canonical
and grand-canonical ensembles of long-range interacting systems
\cite{Misawa2006}. It has been suggested that differences between
ensembles in these systems are usually manifested in the
vicinity of first order phase transitions
\cite{ Barre2002,Ellis2004,Touchette2004,Bouchet2005}.
For recent reviews of systems with long-range interactions see e.g. \cite{Dauxois2002,Campa2007,Mukamel2008,Campa2009,Dauxois2010}.
Finding similar features in nonequilibrium driven systems would
thus provide a further link between the two classes of models.

A particularly interesting model within which the interplay between
drive and long-range interaction may be conveniently analyzed is the
ABC model.  This is a one-dimensional driven model on a ring where
three species of hard-core particles evolve under particle conserving local
stochastic dynamics. The model has been introduced by
Evans et al. \cite{Evans1998,Evans1998b} who observed that although
the dynamics of the model is local, it exhibits long-range order
characterized by phase separation of the three species. It has been
found that in the special case where the average densities of the
three species are equal, the steady state distribution of the ABC
model obeys detailed balance with respect to an effective
Hamiltonian with long-range interactions
\cite{Evans1998,Evans1998b}. This observation suggests that although
detailed balance is not satisfied for unequal densities and the
steady state cannot be expressed in terms of an effective
Hamiltonian, some characteristic features of the long-range
interactions may still be observed in the driven regime where the
densities are unequal. The ABC model has been considered by Clincy
et al. \cite{Clincy2003} in the weak drive regime, where the driving
force scales as the inverse of the system size. In this limit the
model has been found to exhibit a second order phase transition from
the homogeneous state to the phase-separated state at some value of
the drive.

In this paper, we introduce a generalization of the ABC model
allowing for vacancies and processes which do not conserve the
number of particles. We demonstrate that even in the presence of
vacancies and nonconserving processes, the model possesses detailed
balance when the average densities of the three species are equal.
In this case, too, the equilibrium steady state may be expressed in
terms of an effective Hamiltonian with long-range interactions.
This enables us to compare the steady states of the conserving and the
nonconserving dynamics  by studying the free energy of the two ensembles.
 Under the equilibrium conditions where the three
densities are equal, the conserving dynamics lead to a steady state
corresponding to the canonical ensemble, while the nonconserving
dynamics results in a grand-canonical equilibrium state. Since in
both cases the effective Hamiltonian is long-ranged, we find that
the two ensembles yield different phase diagrams. A brief account of
this study is given in \cite{Lederhendler2010a}.

The results presented in this paper may be used as a starting point
for studying the nonequilibrium regime, where the densities are
unequal and the steady state cannot be expressed in terms of an
effective Hamiltonian. Such studies of the original, particle conserving,
ABC model in the case of unequal densities have shown that the second
 order transition found for equal densities, persists for unequal densities, and its
position varies continuously with the densities
\cite{Clincy2003,Bodineau2008}.

The paper is organized as follows: The ABC model is defined and its
properties are summarized in Section \ref{sec:TheABCmodel}. In Section
\ref{sec:GeneralizedABCmodel}, the ABC model is generalized to
include vacancies and nonconserving dynamics. The phase diagrams
for the conserving and nonconserving models under effective
equilibrium conditions are derived in Sections
\ref{sec:PDconserving} and \ref{sec:PDnonConserving}, respectively.
In Section \ref{sec:NonlocalDynamics} the model is further modified
to explore some nongeneric features, found in the
nonconserving phase diagram. We end with some concluding remarks in Section
\ref{conclusions}.

\section{The ABC model}
\label{sec:TheABCmodel}

The ABC model is a prototypical model of driven systems exhibiting
long-range order in one dimension \cite{Evans1998,Evans1998b}. It belongs to a wider class of models exhibiting
similar features, such as the one studied by Lahiri et al. \cite{Lahiri1997,Lahiri2000} in the
context of sedimentation processes. The ABC model is composed of three species of
particles, labeled $A,\, B$ and $C$, occupying a
one-dimensional periodic lattice of length $L$. Each site in the lattice
 is occupied by a single particle.
The number of particles of each type is given by $N_{A},\, N_{B}$ and $N_{C}$
respectively, where $N_{A}+N_{B}+N_{C}=L$. The model evolves under
random sequential dynamics that conserves the particle numbers of the three species, whereby a site is chosen at random and
is exchanged with its neighbor in a clockwise direction according to
the following rates:
\begin{equation}
AB \overset{q}{\underset{1}{\rightleftarrows}} BA, \qquad
BC \overset{q}{\underset{1}{\rightleftarrows}} CB, \qquad
CA \overset{q}{\underset{1}{\rightleftarrows}} AC. \label{eq:CanonicalDynamics}
\end{equation}

For $q=1$ the dynamics is symmetric, and thus obeys detailed
balance. In this case the model relaxes to a homogeneous equilibrium
state, in which all particles are evenly distributed throughout the
lattice. On the other hand for $q\neq 1$, the model relaxes to a
nonequilibrium steady state in which the particles phase separate
into three distinct domains. For $q<1$ the domains are arranged
clockwise in the order $AA\ldots ABB\ldots BCC\ldots C$, and
counterclockwise for $q>1$. This is a strongly phase-separated state
in the sense that fluctuations result in broadening of the domain
boundaries to a finite width, leaving the bulk of the three domains
unmixed in the thermodynamic limit. Since the steady states
corresponding to $q>1$ and $q<1$ are simply related by space
inversion symmetry we take $q<1$ throughout the paper.

As a result of the dynamical asymmetry of the model, the steady
state of a finite system generally exhibits nonzero currents of
particles. The net current of, say, the $A$ particles is determined
by the difference between the rate at which an $A$ particle diffuses
to the right through the $B$ domain, $\sim q^{N_{B}}$, and the rate
at which it diffuses to the left through the $C$ domain, $\sim
q^{N_{C}}$, so that
\begin{equation}
J_{A}\sim q^{N_{B}}-q^{N_{C}}. \label{eq:currents}
\end{equation}
All currents vanish in the thermodynamic limit where
$L\rightarrow\infty$ with $N_{A}/{L},N_{B}/{L},N_{C}/{L}$ kept
fixed. Equation (\ref{eq:currents}) implies that in the special case of
equal average densities,  $N_{A}=N_{B}=N_{C}$, the steady-state
currents vanish even for finite $L$. Thus, although the system is
driven by asymmetric forces, it reaches a steady state which seems
to have no irreversible currents of particles.

This result suggests that at equal densities, detailed balance may
be satisfied, so that the model evolves into an equilibrium steady
state. Indeed, it has been shown that the model obeys detailed
balance with respect to an effective Hamiltonian
\cite{Evans1998,Evans1998b}. This Hamiltonian possesses
long-range interactions, despite the local nature of its dynamics
 (\ref{eq:CanonicalDynamics}). It is defined in terms of the
microscopic configurations of the model, which consist of the set
$\left\{ X_{i}\right\} =\left\{ A_{i},B_{i},C_{i}\right\} $,
$i=1,\ldots,L$, where
\begin{equation}
X_{i}=\left\{ \begin{array}{cc}
1 & {\rm if\, site\,}i{\rm \, is\, occupied\, by\, an\,}X{\rm \, particle}\\
0 & {\rm otherwise.}\end{array}\right.
\end{equation}
The Hamiltonian is then given by
\begin{equation}
\mathcal{H}\left(\left\{ X_{i}\right\}
\right)=\sum_{i=1}^{L-1}\sum_{j=1}^{L-i}\left(A_{i}C_{i+j}+B_{i}A_{i+j}
+C_{i}B_{i+j}\right).\label{eq:CanonicalH}
\end{equation}
The interaction between particles in $\mathcal{H}$ is long ranged,
mean-field like. The total energy of the model scales super-linearly
with its length ($\mathcal{H}\sim L^{2}$) which is
characteristic of systems with long-range interactions.
The periodic boundary conditions imply that the model is
translationally invariant. In Eq. (\ref{eq:CanonicalH}), however,
site $1$ is arbitrarily chosen. One can check that the Hamiltonian
indeed yields the same energy regardless of this choice,
as long as the densities of the three species are equal.
A manifestly translationally invariant form is obtained
by averaging over all possible choices of site $1$, leading to
\begin{equation}
\mathcal{H}\left(\left\{ X_{i}\right\}
\right)=\sum_{i=1}^{L}\sum_{k=1}^{L-1}\frac{k}{L}
\left(A_{i}B_{i+k}+B_{i}C_{i+k}+C_{i}A_{i+k}\right), \label{eq:trans_invariant_H}
\end{equation}
with the periodic boundary condition, $X_{L+i}\equiv X_{i}$.
The two representations were shown to yield the same energy for every
microscopic configuration \cite{Evans1998b}.
In these representations
the energy of the ground state, where the three species are fully
separated, is $L^2/9$.

Using this Hamiltonian, the steady-state distribution of the ABC
model with $N_{A}=N_{B}=N_{C}=L/3$ is given by:
\begin{equation}
P\left(\left\{ X_{i}\right\}
\right)=\frac{1}{Z_{L}}q^{\mathcal{H}\left(\left \{ X_{i}\right\}
\right)},\label{eq:PXi}
\end{equation}
where $Z_{L}=\sum_{\left\{ X_{i}\right\}
}q^{\mathcal{H}\left(\left\{ X_{i}\right\} \right)}$ is the partition sum.
The fact that detailed balance is satisfied with respect to the
Hamiltonian (\ref{eq:CanonicalH}) can be verified by considering an exchange of two
particles, say $AB\rightarrow BA$. According to Eq.
(\ref{eq:CanonicalH}), the resulting change in $\mathcal{H}$ due to
this exchange is $+1$, whereas the reversed process changes
$\mathcal{H}$ by $-1$. Indeed, the exchange of two neighboring
particles of any different species leads to an energy increment
$\Delta\mathcal{H}=\pm 1$. This, together with the expression
(\ref{eq:PXi}) for the distribution function and the transition
rates (\ref{eq:CanonicalDynamics}) leads to detailed balance.

The model has been shown to exhibit phase separation for any $q\neq
1$, while relaxing to a homogeneous state for $q=1$. In order to
study the phase transition between the two types of equilibrium states,
Clincy et al. \cite{Clincy2003} considered the model with  an $L$-dependent $q$
. Taking the limit of weak asymmetry, where
$q\rightarrow1$ as $L\rightarrow\infty$, the model was found to
relax to one of the phases depending on the rate at which $q$
approaches 1 at large $L$. It has been shown that for equal
densities and a transition rate of the form
$q=\exp\left(-\beta/L\right)$, the steady state in the thermodynamic
limit is homogeneous for $\beta<\beta_{c}$ and inhomogeneous for
$\beta>\beta_{c}$ with a second order transition at
$\beta_{c}=2\pi\sqrt{3}$. Since
$P\left(\left\{ X_{i}\right\}
\right)\sim e^{-\beta\mathcal{H}\left(\left \{ X_{i}\right\}
\right)/L}$,
the parameter $\beta$ can be regarded as
the inverse temperature of the model and $1/\beta_c$ as the critical
temperature.

The ABC model has also been studied on an open interval by Ayyer et
al. \cite{Ayyer2009}. In this case the model exhibits detailed
balance for arbitrary average densities of the three species, and
the phase diagram of the model in the entire space of densities has
been derived in the weak asymmetry limit. The mean field approximation
has been shown to be exact in the thermodynamic limit
\cite{Ayyer2009,Fayolle2004,Fayolle2007,Bodineau2008}, and an
analytic expression for the density profiles in the phase-separated
state has been obtained \cite{Ayyer2009}.

In the following sections we generalize the ABC model on a ring to include
nonconserving processes and analyze the resulting phase diagrams in the
weak asymmetry limit.

\section{Generalized ABC model: vacancies and nonconserving processes}
\label{sec:GeneralizedABCmodel}

We now introduce a generalization of the ABC model, allowing a
comparison of two alternative dynamics: particle-conserving and
particle-nonconserving. We begin by introducing vacancies (labeled
$0$) into the lattice. Thus, each site may be occupied by a particle
of either of the species $A,\, B$ or $C$ or may remain vacant, $0$, with
$N_{A}+N_{B}+N_{C}\equiv N \leq L$. Vacant sites are dynamically neutral, so
that a particle of any species may hop to the left or to the right
into a vacant site with equal probability. Hence, the following rule
is added to the exchange rules in Eq. (\ref{eq:CanonicalDynamics}):
\begin{equation}
X0\overset{1}{\underset{1}{\rightleftarrows}}0X,\label{eq:vacancyexchange}
\end{equation}
where $X=A,B,C$.

We proceed by introducing a nonconserving process, whereby triplets
of neighboring particles are allowed to leave or enter the system in
ordered groups:
\begin{equation}
ABC\overset{p q^{3\mu L}}{\underset{p}{\rightleftarrows}}000,\label{eq:addremoverates}
\end{equation}
where $\mu$ is a chemical potential, taken to be equal for all
three species and $p$ is a parameter whose value does not affect the
steady state in the case where detailed balance is satisfied. This
particular form of the nonconserving process is chosen so that the
equal densities condition, $N_{A}=N_{B}=N_{C}=N/3$, could be
maintained.

In general, this model, consisting of the dynamical rates
(\ref{eq:CanonicalDynamics}),(\ref{eq:vacancyexchange}) and (\ref{eq:addremoverates}),
exhibits nonvanishing currents in the steady state, similarly to the
original $ABC$ model. However, as demonstrated below, for equal
densities the model exhibits detailed balance, reaching an
equilibrium state with a distribution:
\begin{equation}
P\left(\left\{ X_{i}\right\} \right)=\frac{1}{Z_{L}}q^{\mathcal{H}_{GC}\left(\left\{
 X_{i}\right\} \right)},
\end{equation}
governed by the Hamiltonian:
\begin{equation}
\mathcal{H}_{GC}\left(\left\{ X_{i}\right\}\right)= \mathcal{H}\left(\left\{ X_{i}\right\}\right)
-\frac{1}{6}N\left(N-1\right)-\mu NL.\label{eq:H_ABCgen}
\end{equation}
Here $ \mathcal{H}\left(\left\{ X_{i}\right\}\right)$ is the
Hamiltonian of the standard ABC model, as given in Eqs. (\ref{eq:CanonicalH}) or
(\ref{eq:trans_invariant_H}).

 We now verify that
the generalized model indeed obeys detailed balance with respect to
$\mathcal{H}_{GC}$. Under particle-conserving processes
(\ref{eq:CanonicalDynamics}) and (\ref{eq:vacancyexchange}),
detailed balance is maintained due to the fact that vacancies do not
affect the energy of a configuration. For the nonconserving process
(\ref{eq:addremoverates}), detailed balance is verified by noting
that the energy of a configuration is invariant under translation of $ABC$
triplets. Namely, $E(\ldots YABC \ldots)=E(\ldots ABCY \ldots)$,
where $E$ is the energy and $Y$ stands for either a particle of any
species or a vacancy. Thus, the change in energy due to depositing
or evaporating $ABC$ triplets is independent of where on the lattice
this process takes place. Depositing a triplet of $ABC$ into a $000$
triplet changes the total energy (\ref{eq:H_ABCgen}) of a system
with an initial particle number $N$ by
\begin{equation}
\Delta\mathcal{H}_{GC}=N+1-\frac{1}{6}\left(6N+6\right)-3\mu
L=-3\mu L,\label{eq:deltaHABC}
\end{equation}
which is consistent with the local dynamical rates in Eq. (\ref{eq:addremoverates}).
Therefore, detailed balance is maintained for
the nonconserving process as well. Equation (\ref{eq:deltaHABC}) implies that the dynamical
parameter $\mu$ is in fact the conjugate field of $N$.

We proceed by considering two cases. The first is a {\it conserving
model} whose dynamical rules consist of Eqs.
(\ref{eq:CanonicalDynamics}) and (\ref{eq:vacancyexchange}). The
second is a {\it nonconserving model} that evolves by all three
types of processes given in Eqs. (\ref{eq:CanonicalDynamics}),
(\ref{eq:vacancyexchange}) and (\ref{eq:addremoverates}). With the
effective Hamiltonian, $\mathcal{H}_{GC}$, the two types of
dynamics correspond to the canonical and the grand canonical
descriptions of the ABC model, respectively.

\section{Phase diagram of the ABC model with conserving dynamics}

\label{sec:PDconserving} We consider the generalized ABC model under
conserving dynamics, in the case of equal densities and in the weak
asymmetry limit, $q=e^{-\beta/L}$. Previous studies of the standard
ABC model for equal densities \cite{Clincy2003} found a second order
phase transition from a homogeneous to an ordered phase at $\beta_c
= 2\pi\sqrt{3}$. This result can be easily extended to the
generalized model with conserving dynamics by noting that the
vacancies do not contribute to the energy of the model, and
thus they are randomly spread in the lattice in the equilibrium state.
This allows us to map each microscopic configuration of the generalized
model to that of the standard ABC model by removing the vacancies, yielding a
'condensed' system of size $N$.  The mapping may be reversed by
adding to the 'condensed' system the $L-N$ vacancies in all possible
arrangements with equal probability, resulting in a  one-to-many
correspondence. The free energies of system with $L-N$ vacancies and its 'condensed' counterpart,
denoted as $\mathcal{F}(N_A,N_B,N_C,L-N)$ and $\mathcal{F}(N_A,N_B,N_C,0)$, respectively,
 differ only by a shift due to the entropy of the vacancies:

\begin{equation}
\mathcal{F}\left(\frac{N}{3},\frac{N}{3},\frac{N}{3},L-N\right)=
\mathcal{F}\left(\frac{N}{3},\frac{N}{3},\frac{N}{3},0\right)
-\ln{{L}\choose{L-N}}.
\end{equation}
Here and throughout this paper the free energy is rescaled by
$\beta$. The 'condensed' $N$-size system is thus equivalent to the
standard ABC model. Writing
$q=\exp\left(-\beta/L\right)=\exp\left(-\beta r/N \right)$,
 where $r= N/L$, it is evident that the 'condensed' system has an effective
inverse temperature of $\beta r$. It therefore exhibits a phase transition at the critical line
\begin{equation}
\beta_{c}=\frac{2\pi\sqrt{3}}{r},\label{eq:Tc}
\end{equation}
which is also the transition line of the corresponding generalized ABC model of length $L$.
For $r=1$, the model contains no vacancies, and we recover the
transition point of original ABC model.

In order to compare the phase diagrams of the conserving and the
nonconserving dynamics, we plot the phase diagrams in the
$\left(1/\beta,\mu\right)$ plane. The chemical potential of the
conserving model is obtained using the Hamiltonian
\begin{equation}
\mathcal{H}_{C}\left(\left\{ X_{i}\right\}\right)= \mathcal{H}\left(\left\{ X_{i}\right\}\right)
-\frac{1}{6}N\left(N-1\right).\label{eq:H_ABCcan}
\end{equation}
with $\mathcal{H}\left(\left\{ X_{i}\right\}\right)$ given by  Eqs. (\ref{eq:CanonicalH}) or
(\ref{eq:trans_invariant_H}). The last term in $\mathcal{H}_{C}$ is chosen so that the Hamiltonian
 differs from the nonconserving Hamiltonian (\ref{eq:H_ABCgen}) only by the term $-\mu NL$, as required
when defining the canonical and grand-canonical Hamiltonians of a model.
The free energy of the conserving model
 in the homogeneous phase is given in the large $N$ and $L$ limit by
\begin{equation}
\mathcal{F}_h\left(\frac{N}{3},\frac{N}{3},\frac{N}{3},L-N\right)
=N\ln{\left(\frac{N}{3}\right)} +
 \left(L-N\right)\ln{\left(L-N\right)}.
\end{equation}

\begin{figure}
\begin{center}
 \includegraphics[scale=0.7]{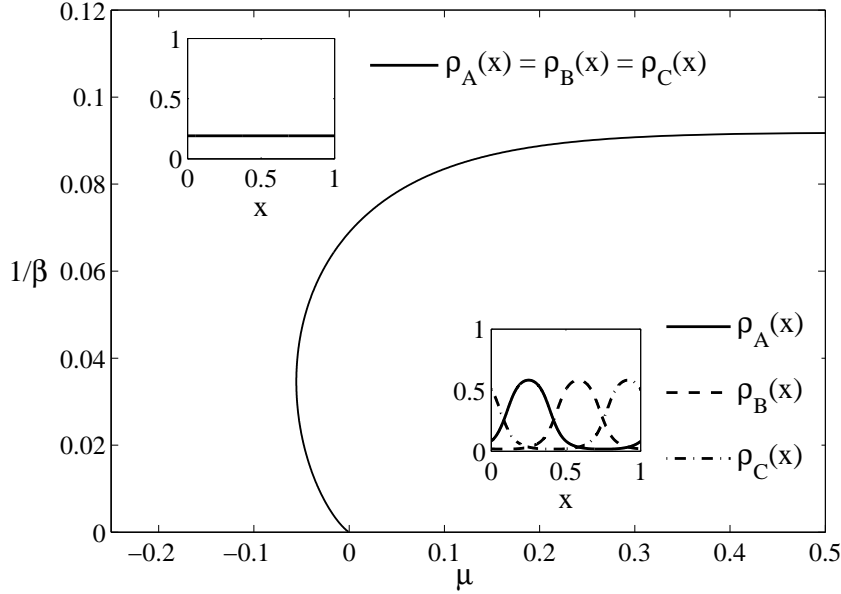}
\end{center}

\caption{\label{fig:PDconserving} The $\left(\mu,\beta^{-1}\right)$ phase
diagram of the generalized ABC model with conserving dynamics,
displaying a second order phase transition line between a
homogeneous and a phase-separated states. Typical density profile,
$\rho_{n}(x)$, in each phase are given in the insets.}

\end{figure}

Due to the specific choice of the constant energy shift in the Hamiltonian
 (\ref{eq:H_ABCcan}), the energy vanishes in the homogeneous state, and only the entropy
 contributes to the free energy.
The chemical potential in the homogeneous phase is thus given by
\begin{equation}
\mu= \frac{1}{\beta}\frac{\partial\mathcal{F}_h}{\partial N}=\frac{1}{\beta} \left[ \ln{\left(\frac{N}{3}\right)} - \ln{\left(L-N\right)} \right],
\end{equation}
and the critical line, where $r_c=2\pi\sqrt{3}/\beta$, can be written as
\begin{equation}
\mu_c=\frac{1}{\beta}\left[\ln\left(\frac{2\pi}{\sqrt{3}\beta}\right)
-\ln\left(1-\frac{2\pi\sqrt{3}}{\beta}\right)\right]. \label{eq:MuC}
\end{equation}
The resulting phase diagram of the conserving model is shown in
Figure \ref{fig:PDconserving}.

We now provide an alternative derivation of this phase diagram by
expanding the free energy of the model in small deviations of the
density profile from the homogeneous solution. This approach yields
more information about the nature of the transition,
and will especially be useful in the next section for analyzing the
nonconserving phase diagram.

For this purpose, we turn to the continuum limit
\cite{Clincy2003,Ayyer2009}, where the local densities of $A,\, B$
and $C$ particles at the point $x=i/L$ are represented by the
density profile $\rho_{n}(x)$, $n=A,B,C$, with
$\rho(x)=\rho_{A}(x)+\rho_{B}(x)+\rho_{C}(x)$. The average particle
density is $r=N/L=\int_{0}^{1}\rho(x)dx$. The steady-state
distribution of the density profiles may be expressed as
$P\left[\rho_{n}(x)\right]=\exp\left\{
-L\mathcal{F}\left[\rho_{n}(x)\right]\right\} $, where
$\mathcal{F}\left[\rho_{n}(x)\right]$ is the free energy functional,
rescaled by $\beta$. The equilibrium profile can thus be found by
minimizing the free energy functional with respect to $\rho_{n}(x)$,
under the equal densities condition,
$\int_{0}^{1}\rho_{A}(x)dx=\int_{0}^{1}\rho_{B}(x)dx=\int_{0}^{1}\rho_{C}(x)dx=r/3$.
The free energy functional for the conserving model is
\begin{eqnarray}
\mathcal{F}[\rho_{n}(x)] & = & \int_{0}^{1}dx\left[\rho_{A}(x)\ln\rho_{A}(x)+\rho_{B}(x)\ln\rho_{B}(x)  \right. \nonumber \\
& + & \left.\rho_{C}(x)\ln\rho_{C}(x)+\left(1-\rho(x)\right)\ln\left(1-\rho(x)\right)\right]  \nonumber \\
& + & \beta\left\{
\int_{0}^{1}dx\int_{0}^{1}dz\left[\rho_{A}(x)\rho_{B}(x+z)+\rho_{B}(x)\rho_{C}(x+z)\right.\right.\nonumber\\
& + & \left.\left.\rho_{C}(x)\rho_{A}(x+z)\right]z-\frac{1}{6}r^{2}\right\}
 ,\label{eq:Fconserving}
\end{eqnarray}
%\int_{0}^{1}dx\int_{0}^{1-x}dz\left[\rho_{A}(x)\rho_{C}(x+z)+\rho_{B}(x)\rho_{A}(x+z)\right.\right. + \nonumber\\
%\left.\left.\rho_{C}(x)\rho_{B}(x+z)\right]-\frac{1}{6}r^{2}\right\}
%
%
where the first integral corresponds to the entropy and the second
integral corresponds to the continuum limit of the Hamiltonian (\ref{eq:H_ABCcan}).

Some characteristics of the equilibrium profile $\rho_{n}(x)$ can be
determined by the symmetry of the model. Due to the cyclic boundary
conditions, $\rho_{A}(x)$, $\rho_{B}(x)$ and $\rho_{C}(x)$ are
periodic functions with period $1$. In addition, since the Hamiltonian
favors phase separation between the three species, we expect the
density profiles in the equal densities case to satisfy
\begin{equation}
\rho_{B}(x) = \rho_{A}\left(x-\frac{1}{3}\right)\,, \qquad
\rho_{C}(x) = \rho_{A}\left(x+\frac{1}{3}\right).\label{eq:BCshift}
\end{equation}
This assumption is further justified in \ref{sec:AnalyticSolution}.
We now use the translation symmetry of the model and set $x=0$ at
the symmetry axis of $\rho_{A}(x)$. The coarse grained state of the
model can thus be represented by a Fourier series for $\rho_A(x)$
\begin{eqnarray}
\rho_{A}(x) & = & \frac{r}{3}+\sum_{m=1}^{\infty}a_{m}\cos\left(2m\pi x\right).
\end{eqnarray}
The relation (\ref{eq:BCshift}) implies that the other two profiles
are given by
\begin{eqnarray}
\rho_{B}(x) & = & \frac{r}{3}+\sum_{m=1}^{\infty}a_{m}\cos\left[2m\pi
\left(x-\frac{1}{3}\right)\right], \nonumber \\
\rho_{C}(x) & = & \frac{r}{3}+\sum_{m=1}^{\infty}a_{m}\cos\left[2m\pi
\left(x+\frac{1}{3}\right)\right]. \label{BCprofiles}
\end{eqnarray}
At high temperatures, $T=1/\beta$, all coefficients $a_{m}$ vanish,
and the profile is homogeneous, with
$\rho_{A}(x)=\rho_{B}(x)=\rho_{C}(x)=r/3$.

In order to find the transition line between the disordered and
ordered phases we expand $\mathcal{F}$ close to the homogeneous
profile in terms of a small perturbation by assuming $a_m\ll 1$.
The amplitudes evolve by $\frac{d a_m}{dt}=-\frac{\partial \mathcal{F}}{\partial a_m}$.
In \ref{sec:ExpandF} we show that the transition to the inhomogeneous phase takes place
 at $\beta=2\pi\sqrt{3}/r$ when the first mode, $a_1$, becomes unstable, whereas all
higher order modes are linearly stable.
Just below this critical line
the higher order modes ($m>1$) are driven by $a_1$ and may be
represented by a power series of $a_1$. The amplitude of $a_1$ may
thus serve as the order parameter of the transition.

The amplitudes of the higher order modes ($m>1$) are obtained by setting
$\partial\mathcal{F}/\partial a_{m}=0$, which yields
to lowest order $a_m\sim a_1^m$ (see \ref{sec:ExpandF}).
The fact that the vacancies are homogeneously distributed in the
 equilibrium state implies that the total particle density is
constant in space, $\rho_0(x)=1-\rho(x)=1-r$, and hence $\delta\rho_{A}(x)+\delta\rho_{B}(x)+\delta\rho_{C}(x)=0$.
From this it follows that all $a_{3m}$ coefficients vanish for $m
\geq 1$. Consequently, the expansion of $\mathcal{F}[\rho_{n}(x)]$ in
powers of $a_1$ is greatly simplified. Up to order $a_1^4$ it
requires terms that involve only the coefficients $a_{1}$ and $a_2$, yielding
\begin{eqnarray}
\mathcal{F}\left[r/3+\delta\rho_{n}(x)\right]=
\mathcal{F}_{{\rm h}}(r)+\left(\frac{9}{4r}-\frac{3\sqrt{3}\beta}{8\pi}\right)a_{1}^{2}\nonumber\\
+\left(\frac{9}{4r}+\frac{3\sqrt{3}\beta}{16\pi}\right)a_{2}^{2}-\frac{27}{8r^{2}}a_{1}^{2}a_{2}
+\frac{81}{32r^{3}}a_{1}^{4}+\mathcal{O}\left(a_{1}^{6}\right),\label{eq:Fexpand}
\end{eqnarray}
where $\mathcal{F}_{{\rm h}}(r)$ is the free energy of the homogeneous
profile given by
\begin{equation}
\mathcal{F}_{{\rm h}}\left(r\right)=r\ln\left(\frac{r}{3}\right)+\left(1-r\right)\ln\left(1-r\right).
\end{equation}
Details of this derivation are given in \ref{sec:ExpandF}.

We can express $a_{2}$ in terms of $a_{1}$
using the equation $\partial\mathcal{F}/\partial a_{2}=0$, which yields:
\begin{equation}
a_{2}=\frac{9\pi}{r\left(\sqrt{3}\beta r+12\pi\right)}a_{1}^{2}.
\end{equation}
We finally obtain the following Landau expansion of the model,
given by the power series of $\mathcal{F}$ in the order parameter $a_1$:
\begin{equation}
\mathcal{F}\left[\rho_{n}(x)\right]=\mathcal{F}_h(r)
+f_{2}a_{1}^{2}+f_{4}a_{1}^{4}+\mathcal{O}\left(a_{1}^{6}\right),
\end{equation}
where
\begin{eqnarray}
f_{2}\left(\beta,r\right) = \frac{9}{4r}-\frac{3\sqrt{3}\beta}{8\pi}, \quad
f_{4}\left(\beta,r\right) = \frac{81}{32r^{3}}\left(\frac{\sqrt{3}\beta
r+6\pi}{\sqrt{3}\beta r+12\pi}\right)>0.\label{eq:fterms}
\end{eqnarray}
By setting $f_{2}=0$ we obtain the same critical line as
in Eq. (\ref{eq:Tc}), shown in Figure \ref{fig:PDconserving}.
The fact that $f_4$ is positive for any value of $r$
indicates that this is a second order transition line,
 from the disordered phase, where $f_2>0$ and
$\mathcal{F}$ is minimized by $a_1=0$, to the ordered phase, where
$f_2<0$, in which $\mathcal{F}$ is minimized by a nonvanishing
$a_1$.

\section{Phase diagram of the ABC model with nonconserving dynamics}

\label{sec:PDnonConserving}

\subsection{The second order line}

\label{sec:NonconservingSecondOrder}

The free energy functional corresponding to the generalized ABC model
with nonconserving dynamics is
\begin{equation}
\mathcal{G}[\rho_{n}(x)]=\mathcal{F}\left[\rho_{n}(x)\right]-\beta\mu
r,
\end{equation}
where $\mathcal{F}\left[\rho_{n}(x)\right]$ is given by Eq.
(\ref{eq:Fconserving}). In order to find the transition between the
disordered and the ordered phases, $\mathcal{G}$ may be expanded
close to the homogeneous profile, as was done in the previous
section in analyzing $\mathcal{F}$. Here, however, in addition to
the modulation of the density profiles of $A,B$ and $C$,
parameterized by $a_{m}$, deviations of the overall density, $r$,
represented by $\delta r$, must be taken into account. Thus, the
$A$-particle density profile close to the transition can be written
as
\begin{equation}
\rho_{A}(x)=\frac{r}{3}+\frac{\delta
r}{3}+\sum_{m=1}^{\infty}a_{m}\cos(2\pi mx),
\label{eq:free_energy_nc}
\end{equation}
where here, again, $\rho_B(x)=\rho_A(x-1/3)$ and $\rho_C(x)=\rho_A(x+1/3)$.
Similarly to the conserving model, the flat
profile of the vacancies implies that all $a_{3m}=0$ for $m \geq 1$.
One can show that to lowest order $a_m\sim a_1^m$ and $\delta r \sim
a_1^2$. This simplifies the expansion of $\mathcal{G}$ considerably.
We carried out the expansion to eighth order in $a_1$, however in order to avoid
lengthy expressions, we outline it here and in \ref{sec:ExpandG} only
to sixth order. In  \ref{sec:ExpandG} we find that

\begin{eqnarray}
\label{eq:G-expand}
\mathcal{G}\left[\rho_{n}(x)\right] & = & \mathcal{G}_{{\rm h}} \left(r\right)+\left(\frac{9}{4r}-\frac{3\sqrt{3}\beta}{8\pi}\right)a_{1}^{2}
 +\left(\frac{9}{4r}+\frac{3\sqrt{3}\beta}{16\pi}\right)a_{2}^{2} \nonumber \\
& - & \frac{9}{4r^{2}}a_{1}^{2}\delta r
  + \left(\frac{1}{2\left(1-r\right)}+\frac{1}{2r} \right)
\left(\delta r\right)^{2}-\frac{27}{8r^{2}}a_{1}^{2}a_{2} \nonumber \\
& + & \frac{81}{32r^{3}}a_{1}^{4}+\frac{243}{32r^{5}}a_{1}^{6}
-\frac{9}{4r^{2}}a_{2}^{2}\delta r +\frac{81}{8r^{3}}a_{1}^{2}a_{2}^{2} \\
& + & \frac{9}{4r^{3}}a_{1}^{2}\delta r^{2}-\frac{243}{32r^{4}}a_{1}^{4}\delta r
-\frac{243}{16r^{4}}a_{1}^{4}a_{2} \nonumber \\
& + & \frac{27}{4r^{3}}a_{1}^{2}a_{2}\delta r +\left(\frac{1}{6\left(1-r\right)^{2}}-
\frac{1}{6r^{2}}\right)\left(\delta r\right)^{3}
+\mathcal{O}\left(a_{1}^{8}\right) \nonumber ,
\end{eqnarray}
where
\begin{equation}
\mathcal{G}_{{\rm h}}\left(r\right)=
r\ln\left(\frac{r}{3}\right)+\left(1-r\right)\ln\left(1-r\right)-\beta\mu r \label{eq:Gh}
\end{equation}
is the free energy of the homogeneous profile.

In order to characterize the nature of the transition
line of the nonconserving model, the expansion in Eq.
(\ref{eq:G-expand}) has to be continued to eighth order in $a_{1}$,
by taking into account terms that involve only the amplitudes $a_1$,$a_2$, $a_4$ and $\delta r$.
The amplitudes are substituted by the following series in $a_1$:
\begin{eqnarray}
\delta r = A_{0,2}a^{2}_1+A_{0,4}a^{4}_1+A_{0,6}a^{6}_1 \nonumber \\
a_{2} = A_{2,2}a^{2}_1+A_{2,4}a^{4}_1+A_{2,6}a^{6}_1  \nonumber \\
 a_{4} = A_{4,4}a^{4}_1 .\label{eq:coeffsCDE_1}
\end{eqnarray}
The coefficients $\{A_{i,j}\}$ are derived from the equilibrium
condition $\partial\mathcal{G}/\partial\left(\delta r\right)=0$ and
$\partial\mathcal{G}/\partial a_{m}=0$ for $m>1$ (see \ref{sec:ExpandG}). Thus, the Landau expansion of $\mathcal{G}$ in
terms of $a_1$ is obtained:
\begin{equation}
\mathcal{G}\left[\rho_{n}(x)\right]=\mathcal{G}\left(r\right)+
g_{2}a^{2}_1+g_{4}a^{4}_1+g_{6}a^{6}_1+g_{8}a^{8}_1\ldots
\label{eq:GExpand}
\end{equation}
The second order coefficient
\begin{equation}
g_{2}\left(\beta,r\right)=f_{2}\left(\beta,r\right)=\frac{9}{4r}-\frac{3\sqrt{3}\beta}{8\pi},
\end{equation}
vanishes at $\beta_{c}=2\pi\sqrt{3}/r$. On the critical line, $\beta=\beta_{c}$,
the fourth order coefficient is:
\begin{equation}
g_{4}\left(\beta_{c},r\right)=\frac{27}{32r^{3}}\left(3r-1\right)~.
\end{equation}
It is positive for $r>1/3$ and it becomes negative for $r<1/3$.
Therefore, there is a multicritical point (MCP) at $r_{{\rm MCP}}=1/3$, with
\begin{eqnarray}
\beta_{{\rm MCP}} = 6 \pi\sqrt{3} \simeq 32.648
 &  \qquad  & \mu_{{\rm MCP}} = -\frac{\ln\left(6\right)}{6\pi\sqrt{3}}\simeq -0.0549 .
\end{eqnarray}
Calculating the higher order coefficients, we find that on the critical line
\begin{eqnarray}
g_{6}\left(\beta_{c},r\right)=\frac{81}{64r^{5}}\left(6r^{2}-5r+1\right),\\
g_{8}\left(\beta_{c},r\right)=\frac{27}{1024r^{7}}\left(1215r^{3}-1692r^{2}+762r-109\right).
\end{eqnarray}

\begin{figure}
\begin{center}
 \includegraphics[scale=0.65]{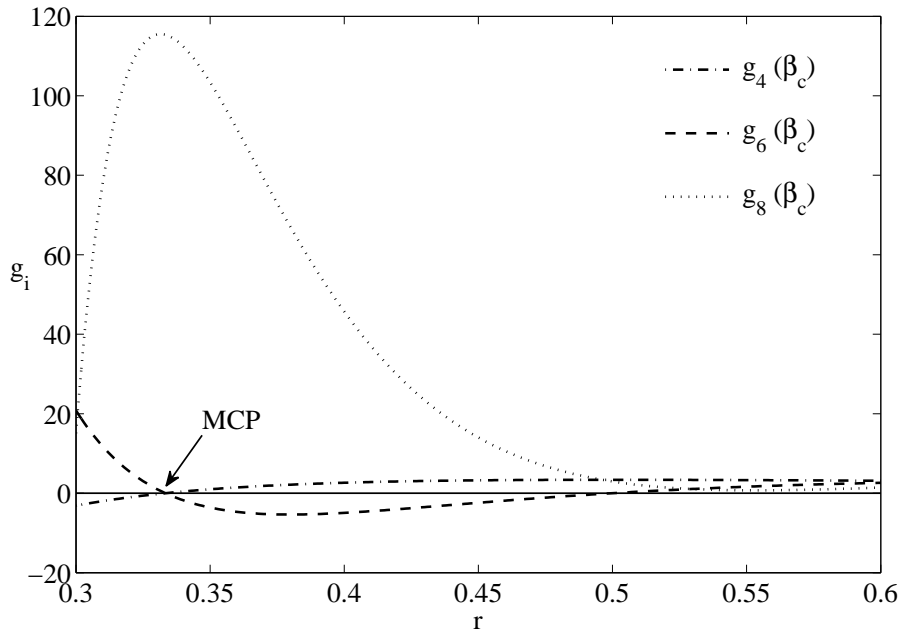}
\end{center}

\caption{\label{fig:gisAtTc}The 4$^{th}$,6$^{th}$ and 8$^{th}$ order coefficients in the
expansion of $\mathcal{G}$ along the critical line. Both the 4$^{th}$ and 8$^{th}$ order
coefficients are positive above the MCP. The 6$^{th}$ order coefficient
is negative just above the MCP, and vanishes at the MCP itself.}

\end{figure}

The sixth order coefficient vanishes at the MCP,
$g_{6}\left(\beta_c,r_{MCP}\right)=0$. This seems to be an
accidental coincidence, and it will be discussed in more detail in
 Section \ref{sec:NonlocalDynamics}. The coefficient becomes
negative just above the MCP. However, the eighth order
coefficient is positive, $g_{8}\left(\beta_c,r_{MCP}\right)>0$, and
large enough so that the second order transition is stable above the
MCP. Thus, the MCP is, in fact, a fourth order critical point. A
plot of the coefficients $g_{4},g_{6}$ and $g_{8}$ along the
critical line as a function of $r$ is shown in Figure
\ref{fig:gisAtTc}. At low densities, $r<r_{MCP}=1/3$,
$g_{4}\left(\beta_{c},r\right)$ is negative and we expect the transition to become
first order.

\subsection{The first order line}

\label{sec:FirstOrderLine}

In this section we complete the phase diagram of the nonconserving
model by evaluating the first order transition below the
multi-critical point. We begin by considering the behavior of the
model at $1/\beta=T=0$, where the density profile has a trivial form.
We then derive the first order line using an analytic expression of
the mean-field profile. Finally we provide a simple approximation
for the first order line at low temperatures, which also yields a
lower bound for the transition for arbitrary temperature.

In the limit $1/\beta=T\to 0$, the entropy may be neglected and the
free energy functional, $\mathcal{G}[\rho_{n}(x)]$, is given by the
energy:
\begin{eqnarray}
\mathcal{G}_{T=0}\left[\rho_{n}(x)\right]=\beta\left\{ \int_{0}^{1}dx\int_{0}^{1}dz\left[\rho_{A}(x)\rho_{B}(x+z)+\rho_{B}(x)\rho_{C}(x+z)\right.\right. \nonumber \\
\left.\left.+\rho_{C}(x)\rho_{A}(x+z)\right]z-\frac{1}{6}r^{2}-\mu r\right\} .
\end{eqnarray}
It is straightforward to verify that the ground state profile is the
fully separated state,
\begin{eqnarray}
\rho_{A}^{{\rm sep}}(x) & = & \left\{ \begin{array}{cc}
r & x<\frac{1}{3}\\
0 & {\rm otherwise}\end{array}\right.,\label{eq:rhosep}
\end{eqnarray}
with $\rho_{B}^{{\rm sep}}(x)=\rho_{A}^{{\rm sep}}(x-1/3)$ and $\rho_{C}^{{\rm sep}}(x)=\rho_{A}^{{\rm sep}}(x+1/3)$.
The free energy of this state,
\begin{equation}
\mathcal{G}_{T=0}(r)=-\beta\left(\frac{1}{18}r^{2}+\mu r\right),
\end{equation}
is minimal at $r=0$ for $\mu<-\frac{1}{18}$ and at $r=1$ for
$\mu>-\frac{1}{18}$. Consequently, there is a discontinuous
transition from an empty system to a fully occupied phase-separated
one at $\mu=-\frac{1}{18}$ and $T=0$. This suggests that there is a
first order transition line, denoted here as
$(T^{\star},\mu^{\star})$, which connects $(0,-\frac{1}{18})$ and
the MCP.

The derivation of the first order transition line requires
 an explicit expression for the density profile, $\rho_{n}(x)$,
 at finite temperatures.
To this end we first compute the profile of the conserving model
using its mapping to the standard ABC
model presented in the first paragraph of Section \ref{sec:PDconserving}. According to this
mapping, the steady state of a conserving model of size $L$
can be extracted from that of the standard ABC model (without
vacancies) of size $N$ with an effective inverse temperature of
$\beta r$. For the standard ABC model we can apply the analytic
solution of the mean-field equations which has been suggested by
Fayolle et al. \cite{Fayolle2004} and derived explicitly by Ayyer et
al. \cite{Ayyer2009} (see also \ref{sec:AnalyticSolution}). The
solution has been formulated for the ABC model on an interval, but
for the case of equal densities it applies also for periodic
boundary conditions. We use it to obtain the profile at an inverse
temperature $\beta r$, and map it back to the profile of the corresponding
generalized model (with vacancies) by multiplying it by $r$, yielding :
\begin{equation}
\label{eq:analytic_profile}
\rho_A(x)=r\frac{1+\mathrm{sn}\left(2\beta r x / \varkappa,k \right) }{\alpha_+ - \alpha_- \mathrm{sn}\left(2\beta r x / \varkappa,k \right)}\quad,
\end{equation}
where $\mathrm{sn}$ stands for the Jacobi elliptic function, and
$\varkappa,\alpha_{+},\alpha_{-},k$ are functions of the parameter
$\beta r$ whose form is given in \ref{sec:AnalyticSolution}. The
profiles for $B$ and $C$ are again given as
$\rho_B(x)=\rho_A(x-1/3)$ and $\rho_C(x)=\rho_A(x+1/3)$.

As shown in \ref{sec:AnalyticSolution} the density profile in Eq. (\ref{eq:analytic_profile})
is also a stationary solution of the mean-field equations of the nonconserving model. These mean-field
equations include, however, an additional constraint,
\begin{equation}
\rho_0^3\left(x\right)=e^{-3\beta\mu}\rho_A\left(x\right)\rho_B\left(x\right)\rho_C\left(x\right),
\end{equation}
which results from the detailed balance condition relating the
 evaporation and deposition processes (\ref{eq:addremoverates}).
This constraint yields the relation between  $\mu,r$ and $\beta$ given by
\begin{equation}
\label{eq:analytic_mu}
\mu=\frac{1}{\beta}\ln{\left[\frac{rK^{1/3}\left(\beta r\right)}{1-r}\right]},
\end{equation}
where $K$ is independent of $x$, and obeys
$Kr^3=\rho_A(x)\rho_B(x)\rho_C(x)$. The dependence of $K$ on the parameter
$\beta r$ is given in \ref{sec:AnalyticSolution}. Equation
(\ref{eq:analytic_mu}) also defines the chemical potential in the
conserving model, where each steady-state profile is also a
stationary solution of the nonconserving model with that value of
$\mu$.

\begin{figure}
\begin{center}
\includegraphics[scale=0.65]{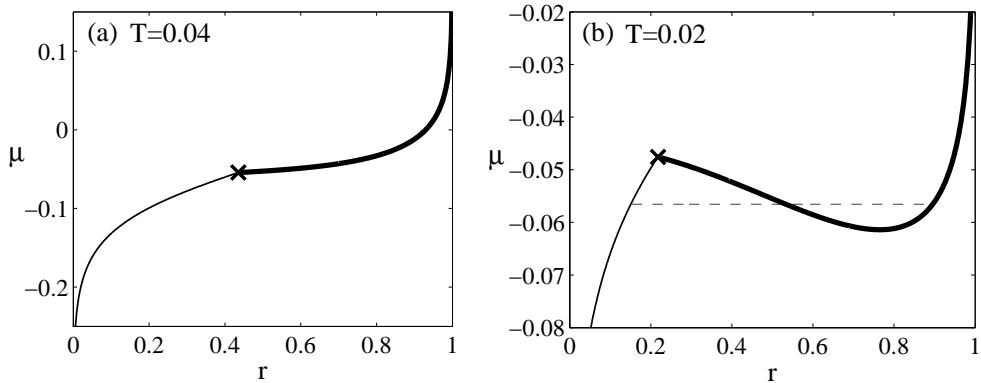}
\end{center}

\caption{\label{fig:MuVersRho} The $\mu(r)$ curve at constant
temperature as calculated in the conserving model using the analytic
solution of the mean-field equations. The thin and thick lines
correspond to the homogeneous and ordered phases, respectively, and
the transition point is marked by $\bf x$. In (a) $T=0.04>T_{MCP}$,
while in (b) $T=0.02<T_{MCP}$. The dashed line in (b) marks the
point where the free energy of the two phases is equal and the
nonconserving model exhibits a first order transition.}
\end{figure}

The resulting $\mu\left(r\right)$ curves for fixed $\beta$ are shown
in Figure \ref{fig:MuVersRho}. The key feature is the region of
$\mu$ in Figure \ref{fig:MuVersRho}b where there are 3 available
solutions for $r$. The solution with the intermediate values of $r$ has negative
compressibility and it is therefore unstable under the nonconserving
dynamics. At the value of $\mu$ for which the two other stable solutions
have the same free energy (denoted by a dashed line) the nonconserving model
undergoes a first order phase transition. The transition point, $\mu^\star$, is
found by evaluating the free energy of the nonconserving model,
$\mathcal{G}$, using the chemical potential and the density profile given above.
In principle, one can deduce the entire phase diagram using this
nonperturbative approach. However, the expansion of the free energy
in Section \ref{sec:NonconservingSecondOrder} is more convenient for
characterizing the nature of the transition.

\begin{figure}
\begin{center}
\includegraphics[scale=0.65]{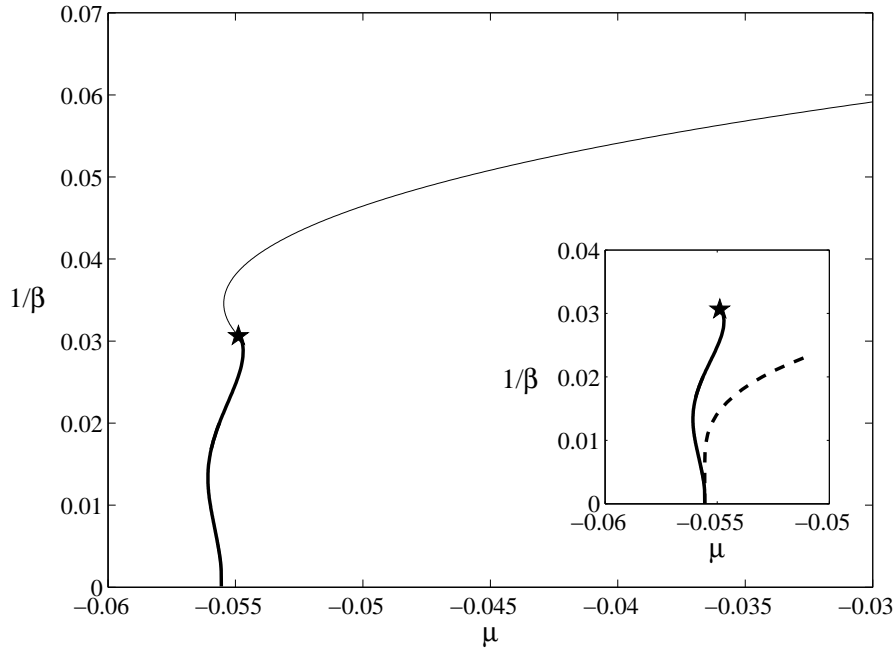}
\end{center}

\caption{\label{fig:PDnonconserving}Phase diagram for the
generalized ABC model with nonconserving dynamics. The second order
(thin line) and first order (thick line) transitions are separated
by a multicritical point ($\star$). The first order line terminates at
($T=0,\mu=-1/18$). The inset displays the lower bound for the first
order line (dashed line) in comparison with the exact first order
transition (see text).}
\end{figure}

The phase diagram of the nonconserving model with the resulting
first order line is presented in Figure \ref{fig:PDnonconserving}.
The numerical evaluation of Eq. (\ref{eq:analytic_profile}) and
(\ref{eq:analytic_mu}) requires numerical precision that grows
linearly with $\beta=1/T$ making it prohibitive at low temperatures.
To study this limit we now present a simple approximation for the
transition line which also yields a lower bound for the transition
at all temperatures.

We first observe that above the transition curve
($T>T^{\star}$ or $\mu<\mu^{\star}$) the optimal profile is
homogeneous, with $\rho_{n}(x)=r/3$. The value of the average
density $r$ is determined by minimizing the free energy of the
homogeneous profile (\ref{eq:Gh})
with respect to $r$. As the model approaches the transition point,
the free energy, $\mathcal{G}$, develops a local minimum
corresponding to a inhomogeneous density profile. The transition
temperature may be defined as the lowest temperature for which the
free energy of the homogeneous profile is lower than that of any
other profile. Instead of considering any possible profile, we look
at the totally separated profile in Eq. (\ref{eq:rhosep}), whose
corresponding free energy is given by
\begin{equation}
\mathcal{G}_{{\rm sep}}(r)=r\ln\left(r\right)+\left(1-r\right)\ln\left(1-r\right)-
\beta\left(\frac{1}{18}r^{2}+\mu r\right).
\end{equation}
For $T>0$ the model eventually displays an ordered phase that is not
fully separated and therefore has a lower free energy (due to
entropic effects) than $\mathcal{G}_{{\rm sep}}(r)$. Hence, for a
given $\mu$ the temperature at which
\begin{equation}
\min_{r}\mathcal{G}_{{\rm h}}(r)=\min_{r}\mathcal{G}_{{\rm sep}}(r),\label{eq:hVSsep}
\end{equation}
is a lower bound for the transition temperature, $T^\star$. At low
temperatures, the first order transition line approaches this lower
bound and the ordered profile is well represented by the fully
phase-separated one. Thus, this bound provides a
good approximation for the behavior of the system at low
temperatures ($T\ll T_{MCP}$), as shown in the inset of Figure
\ref{fig:PDnonconserving}.

\subsection{Monte Carlo simulations}

The picture emerging from the continuum limit is supported
by the results of Monte Carlo (MC) simulations, performed under
effective equilibrium conditions, $N_{A}=N_{B}=N_{C}=N/3$, with both
conserving and nonconserving dynamics.

In the nonconserving simulations, the state of the system is
updated according to the following procedure: At each time-step a
single site is randomly selected. The type of move to attempt is
also chosen randomly: with probability $1/2$ an attempt is made to
exchange the particle of the chosen site with its right-hand
neighbor and with probability $1/2$ an evaporation or deposition
process is attempted. In the latter case, if the chosen site is
occupied by a $B$ particle with an $A$ to its left and a $C$ to its
right, then the removal of the triplet $ABC$ is attempted. If the
chosen site is vacant, and it is surrounded on both sides by vacant
sites as well, the condensation of an $ABC$ triplet is attempted.
Finally, the move is accepted with probability
\begin{equation}
P=\left\{ \begin{array}{cc}
\exp\left(-\beta \frac{\Delta \mathcal{H}_{GC}}{L}\right) & \Delta \mathcal{H}_{GC}> 0\\
1 & {\rm otherwise}\end{array}\right.,\label{eq:PMCstep}
\end{equation}
where $\Delta \mathcal{H}_{GC}$ is the change in energy due to
the chosen move, as determined by Eq. (\ref{eq:H_ABCgen}) (possible
values are $0$, $\pm 1$ and $\pm 3\mu L$). This procedure leads to a
steady state which obeys detailed balance with respect to the nonconserving Hamiltonian
(\ref{eq:H_ABCgen}).

\begin{figure}
\begin{center}
\includegraphics[scale=0.625]{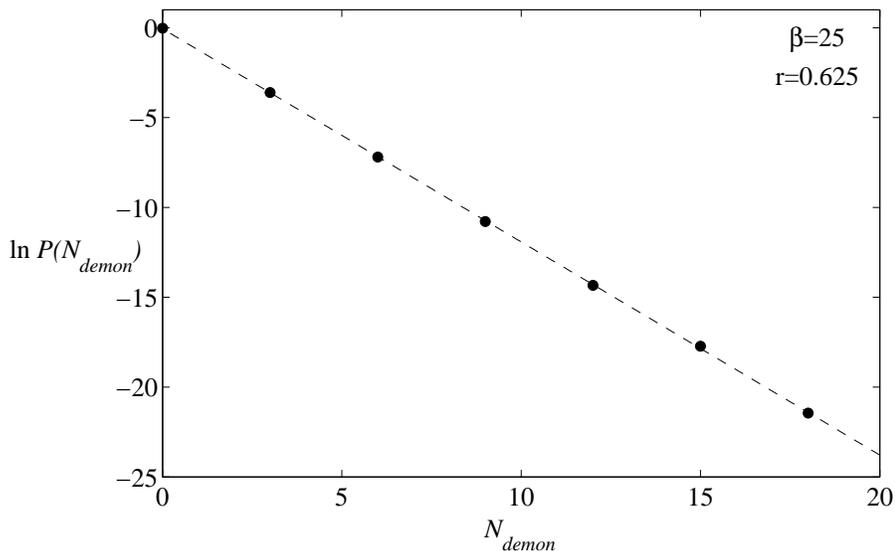}
\end{center}

\caption{\label{fig:PvsN_demon} Natural logarithm of the probability
distribution of the 'demon' occupation number as obtained by MC
simulations of the conserving dynamics for $L=1800$, $N=1125$ and
$\beta=25$. Linear fit (dashed line) of the simulation points yields
$\mu\simeq 0.047$. Such simulations yield the $\mu(r)$ curve.}
\end{figure}

In order to compare the results of these simulations with those of the
conserving simulations one has to calculate the chemical potential,
 $\mu$, for a given temperature under conserving dynamics.
This can be done by employing in the conserving simulations
 a method similar to the Creutz algorithm for
 microcanonical MC simulations \cite{Creutz1983}.
The idea is to perform a constrained nonconserving simulation so
that the average density, $r$, is allowed to fluctuate only below its initial value, $r(0)$,
 while extracting the value of $\mu$ from these fluctuations.
The simulation is executed as in the nonconserving case by selecting a site and an
attempted move. Steps in which neighboring particles are exchanged are
accepted according to Eq. (\ref{eq:PMCstep}). The particle nonconserving
steps are performed in conjunction with an additional single degree of freedom, termed 'demon',
 that exchanges particles with the system.
The 'demon' is initially empty. An attempt of removing $ABC$ triplet is accepted with
 probability $1$, and the removed
particles are added to the 'demon'. Steps that require the
deposition of an $ABC$ triplet on three vacant sites are accepted
only if the 'demon' is not vacant and with a rate given by  Eq. (\ref{eq:PMCstep}) with $\mathcal{H}_{GC}$ replaced by the
canonical Hamiltonian, $\mathcal{H}_C$ (\ref{eq:H_ABCcan}).
The triplet is then removed from the 'demon'.
As a result the average density is allowed to fluctuate, but only to
states with an average density below $r(0)$. Fluctuations
with higher densities are rejected. In the thermodynamic limit this
procedure yields the canonical distribution of the system with
density $r(0)$, even for systems with negative compressibility \cite{Mukamel2005}.
The probability distribution of the number of particles in the 'demon',
 $P\left(N_{demon}\right)$, is recorded during the simulation. The chemical
potential, $\mu$, is determined from this distribution using the
equilibrium relation $P\left(N_{demon}\right)\sim\exp\left(-\beta\mu
N_{demon}\right)$. A typical distribution is given in Figure
\ref{fig:PvsN_demon}, from which $\mu$ is extracted by applying a
linear fit to $\ln\left[P\left(N_{demon}\right)\right]$.

%Under conditions of detailed balance this definition of the chemical
%potential should coincide with the definition in Eq.
%(\ref{eq:MuDef}) used for calculating the first order transition.

%

\begin{figure}[htp]
\centering
\includegraphics[width=1.0\textwidth]{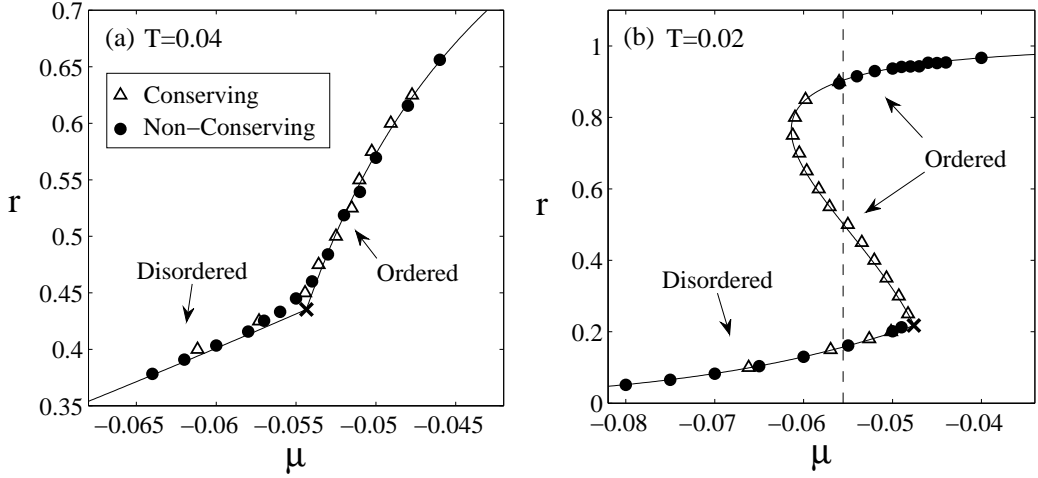}

\caption{\label{fig:rhoVSmu} The $\mu(r)$ curve of the conserving
($\triangle$) and nonconserving ($\bullet$) dynamics obtained by direct
simulations for $L=1800$.
The simulations were performed at two temperatures (a) $T=0.04$,
above the MCP, and (b) $T=0.02$, below the MCP ($T_{{\rm
MCP}}\simeq0.03$). In (a) both the conserving ($\triangle$) and
nonconserving ($\bullet$) simulations result in the same curve
within the numerical accuracy. In (b) the two types of dynamics
yield different curves, with the nonconserving dynamics displaying
a discontinuity in r at a first order transition. Hysteretic
behavior is observed in the nonconserving case. The expected first
order transition point, obtained by minimization of $\mathcal{G}$,
is indicated by the dashed vertical line. The solid lines correspond
to the analytic solution of the mean field equations and so is the {\bf
x}, which denotes the second order transition point in conserving
model.} \hfill

\end{figure}

The simulation results are displayed and compared with the
mean-field solution (\ref{eq:analytic_mu}) in Figure
\ref{fig:rhoVSmu}, where the average density, $r$, is plotted as a
function of the chemical potential, $\mu$. Above the multicritical
point, in Figure \ref{fig:rhoVSmu}a, we see  that the conserving
and nonconserving simulations follow the same curve. The two types
of dynamics are thus equivalent, and undergo a second order
transition at the critical point (\ref{eq:MuC}) marked
 in the figure by {\bf x}. Below
the multicritical point, in Figure \ref{fig:rhoVSmu}b, the
conserving simulation exhibits a similar second order transition,
whereas the nonconserving simulation shows a discontinuity in $r$.
At the intermediate-density states we find negative compressibility
in the conserving simulation. These states are unstable under the
nonconserving dynamics, where $r$ can fluctuate freely. The
nonconserving model thus exhibits a discontinuity in $r$, accompanied
by hysteretic behavior.
This hysteresis is an indication of the first order transition.
The value of $\mu$ at the transition, as found in the thermodynamic
limit using the analytical procedure discussed in Section \ref{sec:FirstOrderLine},
is denoted by the dashed line. This thus provides a direct observation of the
inequivalence of two ensembles. The results of the simulations fit
very well the mean-field solution (\ref{eq:analytic_mu}).

\section{Further generalization of the model}

\label{sec:NonlocalDynamics} The nonconserving ABC model with equal
densities exhibits a seemingly accidental coincidence whereby the
sixth order coefficient in the expansion of the free energy
vanishes at the MCP, i.e. $g_{2}=g_{4}=g_{6}=0$. If indeed this
feature is accidental one expects any modification of the model to
remove this degeneracy. In this section we consider a simple
generalization of the nonconserving dynamics that maintains
detailed balance under the equal densities condition. Thus, the
features of the phase diagram may be found using the expansion of
free energy functionals, as presented in the previous sections.

The generalization consists of replacing the factor $1/6$ in the
Hamiltonian in Eq. (\ref{eq:H_ABCgen}) by a free parameter $\gamma$,
\begin{eqnarray}
\mathcal{H}_{GC,\gamma}\left(\left\{ X_{i}\right\} \right)
=\mathcal{H}\left(\left\{ X_{i}\right\} \right) -\gamma
N\left(N-1\right)-\mu NL, \label{eq:gammaH}
\end{eqnarray}
where $\mathcal{H}\left(\left\{ X_{i}\right\} \right)$ is given by
Eqs. (\ref{eq:CanonicalH}) or  (\ref{eq:trans_invariant_H}) . By imposing the condition for detailed
balance with respect to the distribution $P_{\gamma}\left(\left\{
X_{i}\right\}\right)= q^{\mathcal{H}_{GC,\gamma}\left(\left\{
X_{i}\right\} \right)}/Z_{L}$ for systems with equal densities
$N_{A}=N_{B}=N_{C}$, and reversing the argument of Eq.
(\ref{eq:deltaHABC}), we find that detailed balance is maintained
for the following evaporation and deposition rates:
\begin{equation}
ABC\overset{pq^{\Delta\mathcal{H}_{GC,\gamma}}}{\underset{p}
{\rightleftarrows}}000,\label{eq:gamma_rates1}
\end{equation}
where
\begin{equation}
\Delta\mathcal{H}_{GC,\gamma}=3\mu
L-\left(N+1\right)\left(1-6\gamma\right). \label{eq:gamma_rates2}
\end{equation}
Thus the non conserving dynamics corresponding to the modified Hamiltonian
(\ref{eq:gammaH}) consists of the processes
(\ref{eq:CanonicalDynamics}), (\ref{eq:vacancyexchange}) and
(\ref{eq:gamma_rates1}). For $\gamma \neq 1/6$ the evaporation rate
depends on the particle number, $N$, and therefore the dynamics is
{\it nonlocal}.

We now analyze the phase diagram corresponding to this generalized
model in the limit of weak asymmetry, $q=\exp\left(-\beta/L\right)$.
The continuum-limit calculation of the phase diagram can be
repeated, producing Landau expansions of the free energy functionals
$\mathcal{F}_{\gamma}$ and $\mathcal{G}_{\gamma}$. In the conserving
dynamics the particle number $N$ is a constant, and hence the
$\gamma$-term has no effect on the expansion of
$\mathcal{F}_{\gamma}$. Therefore, regardless of $\gamma$, the
conserving model exhibits a second order transition at
$\beta_{c}=2\pi\sqrt{3}/r$.

The expansion of $\mathcal{G}_{\gamma}$, as detailed in
\ref{sec:ExpandG}, is given by,
\begin{equation}
\mathcal{G}_{\gamma}\left[\rho_{n}(x)\right]=\mathcal{G}_{\gamma}\left(r\right)
+g_{2}^{\gamma}a^{2}_1+g_{4}^{\gamma}a^{4}_1+g_{6}^{\gamma}a^{6}_1+g_{8}^{\gamma}a^{8}_1+\ldots,
\label{eq:GExpand_gamma}
\end{equation}
where $g_{2}^{\gamma}=g_{2}$ is independent of $\gamma$. It vanishes
on the critical line $\beta=\beta_{c}=2\pi\sqrt{3}/r$. On this line
the fourth order coefficient is given by
\begin{eqnarray}
g_{4}^{\gamma}\left(\beta_{c},r\right)=\frac{27}{32r^{3}}\left[\frac{9r-3
+4\pi\sqrt{3}\left(1-6\gamma\right)\left(1-r\right)}{3+2\pi\sqrt{3}\left(1-6\gamma\right)\left(1-r\right)}\right],
\end{eqnarray}
which is positive at all densities satisfying
\begin{equation}
r>r_{{\rm
MCP}}=\frac{\sqrt{3}-4\pi\left(1-6\gamma\right)}{3\sqrt{3}-4\pi\left(1-6\gamma\right)}.
\end{equation}
The coefficient vanishes at $r=r_{MCP}$ which yields:
\begin{eqnarray}
\beta_{{\rm MCP}} & = & \frac{6\pi\left[3\sqrt{3}-4\pi(1-6\gamma)\right]}{3-4\pi\sqrt{3}(1-6\gamma)}\nonumber\\
\mu_{{\rm MCP}} & = & \frac{1}{6}\left[3-4\pi\sqrt{3}(1-6\gamma)\right]\times \nonumber \\
& & \left\{\frac{2(1-6\gamma)}{9-4\pi\sqrt{3}(1-6\gamma)}+\frac{\ln\left[\frac{1}{18}\left(3-4\pi\sqrt{3}(1-6\gamma)\right)\right]}
{\pi\left(3\sqrt{3}-4\pi(1-6\gamma)\right)}\right\}.
\end{eqnarray}
Note that as $\gamma\rightarrow\left(4\pi-\sqrt{3}\right)/24\pi\simeq 0.144$
(from above), the density $r_{{\rm MCP}}$ vanishes, and $g_{4}$ is
positive for all values of $0<r\leq 1$.

\begin{figure}
\begin{center}
\includegraphics[scale=0.62]{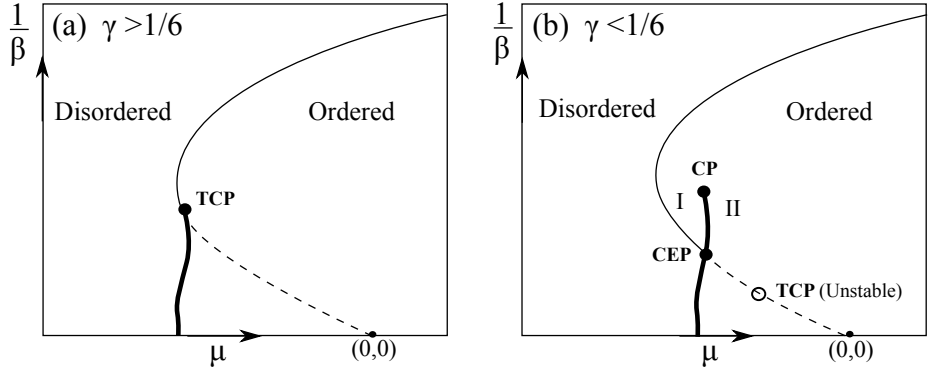}
\end{center}

\caption{\label{fig:SchematicTvsMu}Schematic phase diagrams of the
modified nonconserving ABC model with nonlocal dynamics for (a)
$\gamma\geq\frac{1}{6}$ and (b) $\gamma<\frac{1}{6}$. In both cases
the phase diagram is composed of a second order transition (thin
line) at high temperatures, which becomes first order (thick line)
at low temperatures. The dashed line marks the continuation of the
second order line, which is unstable for the nonconserving
dynamics. The first order line intersects the $T=0$ axis at
$\mu=\frac{1}{9}-\gamma$. In (a) The two lines meet at a tricritical
point (TCP), as in the local dynamics case, $\gamma=\frac{1}{6}$. On the
other hand in (b) the first order transition line intersects the
second order line at a critical end point (CEP). The second order
line terminates at this point while the first order line continues
into the ordered phase, where it marks a transition between two
distinct phase-separated states, with low (I) and high (II)
densities, and terminates at a critical point (CP). The hollow
point on the unstable second order line, denotes the TCP where
$g_2=g_4=0$ and $g_6<0$. This point and the dashed segment of the
second order line are preempted by the first order line and are thus
not accessible within the nonconserving dynamics. For particles
conserving dynamics the phase diagram exhibits a second order line,
composed of the thin and dashed lines in the figure, for any value
of $\gamma$. }
\end{figure}

The phase diagram in the vicinity of the MCP depends strongly on the
sign of the sixth order coefficient, $g_{6}^{\gamma}$, at the MCP
where $g_{2}^{\gamma}=g_{4}^{\gamma}=0$. From the expansion of
$\mathcal{G}_{\gamma}$ we find
\begin{equation}
\left(g_{6}^{\gamma}\right)_{{\rm MCP}}=
\frac{\sqrt{3}\pi\left(1-6\gamma\right)\left[3-2\sqrt{3}\pi\left(1-6\gamma\right)\right]\left[4\sqrt{3}\pi\left(1-6\gamma\right)-9
\right]^{5}}{8\left[3-4\sqrt{3}\pi\left(1-6\gamma\right)\right]^{5}},
\end{equation}
For $\gamma>\frac{1}{6}$, one has $\left(g_{6}^{\gamma}\right)_{{\rm
MCP}}>0$, and the MCP is a tricritical point (TCP). Thus, the phase
diagram in this case consists of a second order line which becomes
first order below the TCP. However for $\gamma<\frac{1}{6}$, the
coefficient $\left(g_{6}^{\gamma}\right)_{{\rm MCP}}$ is negative
and the trictitical point is unstable. As a result  the first order
line intersects the second order line above the TCP. Thus the phase
diagram consists of a critical line which terminates at a first
order line at a critical end point (CEP) where $g_{2}^{\gamma}=0,\,
g_{4}^{\gamma}>0 \,$ and $g_{6}^{\gamma}<0$. This point is located
\textit{above} the TCP (which is unstable). The first order line
continues into the ordered phase, and it ends at a critical point
(CP), as shown schematically in Figure \ref{fig:SchematicTvsMu}.
Above the CEP the first order line marks a transition between two
ordered phases with low (I) and high (II) densities. As $\gamma$
approaches $1/6$ from below, the CEP, CP and TCP approach each
other. They merge at $\gamma=1/6$, yielding a fourth order critical
point.

To complete the analysis of the phase diagram one has to determine
the first order line. This can be done by using the exact solution
for the density profiles, as discussed in Section
\ref{sec:FirstOrderLine}. The location of the first order transition at $T\to 0$
 can be derived by equating the free energies of the homogeneous and the fully
phase-separated states
\begin{equation}
\min_{r}\mathcal{G}_{\gamma,{\rm h}}(r)=\min_{r}\mathcal{G}_{\gamma,{\rm sep}}(r),
\end{equation}
where
\begin{eqnarray}
\mathcal{G}_{\gamma,{\rm h}}(r) =  r\ln\left(\frac{r}{3}\right)+\left(1-r\right)\ln\left(1-r\right)+\beta\left(\frac{1}{6}-\gamma\right)r^{2}-\beta\mu r, \label{eq:Gh_gamma}
\end{eqnarray}
and
\begin{equation}
\mathcal{G}_{\gamma,{\rm sep}}(r) = r\ln\left(r\right)+\left(1-r\right)\ln\left(1-r\right)
+\beta\left(\frac{1}{9}-\gamma\right)r^{2}-\beta\mu r.
\end{equation}
At $T=0$ the model exhibits a
discontinuity in the total density from $r=0$ to $r=1$ at $\mu=1/9-\gamma$,
and thus undergoes a first order transition at that point.

 Figure
\ref{fig:SchematicTvsMu} shows a schematic representation of the
phase diagrams of the generalized ABC model with nonlocal dynamics
for conserving and nonconserving dynamics above and below
$\gamma=1/6$.
As expected, the nongeneric feature of a fourth order critical point
has been removed by slight modification of the model. Generically,
we expect the multicritical point to become either a tricritical point or a
critical end point, as shown in the figure.

\section{Conclusions}
\label{conclusions}

 In this paper we generalized the ABC model to include vacancies
and processes that do not conserve the particle number. This enables us to analyze and
compare the phase diagrams of the conserving and the nonconserving
models. We have shown that in the case where the average densities
of the three species are equal, the dynamics of the generalized
model obeys detailed balance with respect to a Hamiltonian with
long-range interactions, despite the fact that the dynamics is local.
Studying the $\left(\mu,T\right)$ phase diagrams of the model for equal densities,
 we found that in the conserving case it is composed of a second order line
 separating the homogeneous and phase-separated states, while in the
nonconserving case the second order line becomes first order at low temperatures.
 The analysis of the phase diagram has been carried out
by studying the Helmholtz and Gibbs free energies for the
conserving and nonconserving dynamics, respectively, in the
continuum limit. As has been shown in the past this limit yields the
exact steady states in the thermodynamic limit. In this study we
applied a critical expansion of the free energy near the homogeneous
phase and an exact solution of the mean-field equations for the
density profiles of the phase-separated state. The results
of this analysis are verified by direct Monte Carlo simulations
of the two types of dynamics.

The fact that the two types of dynamics result in rather different
phase diagrams can be associated with the inequivalence of the
canonical (conserving) and grand-canonical (nonconserving)
ensembles in systems with long-range interactions. We find that as
expected from studies of long-range interacting systems, the two
ensembles yield different steady states in the region where the
grand canonical ensemble displays a first order transition.

We expect the generalized ABC model to display similar behavior even
for small deviations from the equal densities case, where detailed
balance is not satisfied. The present study can thus serve as a
starting point for a study of the ABC model out of equilibrium. It
could provide an interesting correspondence between some properties
of the well-understood equilibrium systems with long-range
interactions and those of the less well-understood nonequilibrium
driven models. Details of a study of the generalized ABC model with
unequal densities will be published elsewhere \cite{cohen2010}. An
interesting driven model where ensemble-inequivalence has recently
been observed is the zero-range process \cite{Grosskinsky2008}. In
this model, the drive, provided by the spatial asymmetry of the
transition rates, does not influence the steady state. This state
thus remains the same as the steady state of the nondriven
equilibrium model, where the transition rates are symmetric, which
can be expressed in terms of a Hamiltonian. By contrast, the ABC
model provides a framework within which one can readily probe
 nonequilibrium steady states which are not expressed by a Hamiltonian.

\ack We thank Amir Bar, Shamik Gupta, Ori Hirschberg, Yariv Kafri and
Gunter M. Sch\"utz for helpful
discussions. The support of the Israel Science Foundation (ISF) and
the Minerva Foundation with funding from the Federal German Ministry
for Education and Research is gratefully acknowledged.

\appendix

\section{Critical expansion of the conserving free energy}

\label{sec:ExpandF}

We present the critical expansion of the free energy of the conserving model.
In the continumm limit, the free energy, rescaled by $\beta$, is given by
 $\mathcal{F}=\beta \epsilon - s$, where
\begin{eqnarray}
s\left[\rho_{n}(x)\right] & = & -\int_{0}^{1}dx\left[\rho_{A}(x)\ln\left(\rho_{A}(x)\right)+\rho_{B}(x)\ln\left(\rho_{B}(x)\right)\right.\nonumber\\
& & \left.+\rho_{C}(x)\ln\left(\rho_{C}(x)\right)+\left(1-\rho(x)\right)\ln\left(1-\rho(x)\right)\right]
\label{eq:entropyTerm}
\end{eqnarray}
is the entropy per site of the profile, derived from simple combinatorial considerations, and
\begin{eqnarray}
\epsilon \left[\rho_{n}(x)\right]  & = & \int_{0}^{1}dx\int_{0}^{1}dz\left[\rho_{A}(x)\rho_{B}(x+z)+\rho_{B}(x)\rho_{C}(x+z)\right.\nonumber\\
& & \left.+\rho_{C}(x)\rho_{A}(x+z)\right]z-\frac{1}{6}r^{2}\label{eq:enenrgyTerm}
\end{eqnarray}
is the energy per site given continuum limit of the conserving Hamiltonian (\ref{eq:H_ABCcan}),
for $n=A,B$ or $C$.

We begin by expanding the steady-state profile close the homogeneous solution in the most
general form :
\begin{equation}
\rho_{n}(x)=\frac{r}{3}+\delta\rho_{n}(x) =
\frac{r}{3}+\sum_{m=-\infty}^{\infty}\alpha_{n,m}e^{2m\pi ix}.
\end{equation}
For the profile to be real
we assume $\alpha_{n,-m}=\alpha_{n,m}^\star$. The time evolution of
these modes is set by
 $\frac{d \alpha_{n,m} }{ dt}= -\frac{\partial \mathcal{F} }{ \partial \alpha_{n,-m}}$
 which yields to lowest order in $\alpha_{n,m}$:
\begin{equation}
\frac{d}{dt}\vec{\alpha}_{m}=\left[\frac{i\beta}{2\pi m}\left(\begin{array}{ccc}
0 & 1 & -1\\
-1 & 0 & 1\\
1 & -1 & 0\end{array}\right)-\frac{3}{r}I\right]\vec{\alpha}_{m},
\end{equation}
where $\vec{\alpha}_{m}=(\alpha_{A,m},\alpha_{B,m},\alpha_{C,m})$ and $I$ is
the $3\times 3$ unit matrix.
For the $m^{th}$ mode, the highest eigenvalue of the matrix above is
$\frac{\beta \sqrt{3} }{2\pi m}-\frac{3}{r}$ with the eigenvector
$(1,e^{-2\pi i /3},e^{2\pi i /3})$. As the
temperature is decreased ($\beta$ increased) the first mode to become unstable
is $m=1$ (at $\beta=2\pi\sqrt{3}/r$), while the other modes are linearly stable.
Just below this transition line the higher modes ($m>1$) are driven by the
$m=1$ mode. The $m^{th}$ mode is driven to lowest order by a term of the form
$\alpha_{n,1}^m\alpha_{n,-m}$ in the
Taylor expansion of the logarithmic function in the entropy. We thus
obtain to lowest order $\alpha_{n,m}\sim \alpha_{n,1}^m$ with the eigenvector
$(1,e^{-2m\pi i /3},e^{2m\pi i /3})$. We can therefore simplify our expansion and set
\begin{equation}
\rho_{A}(x)=\frac{r}{3}+\delta\rho_{A}(x) = \frac{r}{3}+\sum_{m=1}^{\infty}a_{m}\cos(2m\pi x), \label{eq:rhoA}
\end{equation}
with $\rho_{B}(x)=\rho_{A}(x-1/3)$ and $\rho_{C}(x)=\rho_{A}(x+1/3)$. This latter form
is also justified in the work of Ayyer et al. \cite{Ayyer2009}, who proved that the ordered profile
of the model is unique and obeys this symmetry.

 As explained in Section \ref{sec:PDconserving}, the flat
profile of the vacancies implies that
$\delta\rho_0=-(\delta\rho_{A}+\delta\rho_{B}+\delta\rho_{C})=0$ and
therefore $a_{3m}=0$. We wish to evaluate the free energy up to
order $a_1^4$ and thus expand the entropy and energy only in terms
of $a_1$ and $a_2$.

The entropy of the $A$ particles is given by the first term in the RHS of
 Eq. (\ref{eq:entropyTerm}) :
\begin{eqnarray}
s_A=-\int_{0}^{1}dx\left[\frac{r}{3}+\delta\rho_{A}(x)\right]\ln\left[\frac{r}{3}+\delta\rho_{A}(x)\right] = \nonumber\\
\int_{0}^{1}dx \left\{
\frac{r}{3}\ln\left(\frac{r}{3}\right)+\left[1+\ln\left(\frac{r}{3}\right)\right]\delta\rho_{A}(x)
+\frac{3}{2\rho}\left(\delta\rho_{A}(x)\right)^{2}\right.\nonumber\\
\left.-\frac{3}{2\rho^{2}}\left(\delta\rho_{A}(x)\right)^{3}
+\frac{9}{4\rho^{3}}\left(\delta\rho_{A}(x)\right)^{4} \right\}+\mathcal{O}\left(\delta\rho_{A}(x)\right)^{5}.
\end{eqnarray}
After performing the integral we obtain:
\begin{eqnarray}
-s_{A} & = & \frac{r}{3}\ln\left(\frac{r}{3}\right)+a_{1}^{2}\left(\frac{3}{4r}\right)+a_{2}^{2}\left(\frac{3}{4r}\right) \nonumber \\
& + &a_{1}^{4}\left(\frac{27}{32r^{3}}\right)-a_{1}^{2}a_{2}\left(\frac{9}{8r^{2}}\right).
\end{eqnarray}
Under the condition of equal densities one has $s_{A}=s_{B}=s_{C}$, and
therefore the total entropy of the particles is equal to $3s_{A}$.
The total particle density is constant in space,
$\rho(x)=r$, and thus the entropy of the vacancies, $\left(1-r\right)\ln\left(1-r\right)$,
 contributes a constant term to the expansion.
The total entropy of the system is thus:
\begin{eqnarray}
 -s & = & r\ln\left(\frac{r}{3}\right)+\left(1-r\right)\ln\left(1-r\right)+a_{1}^{2}\left(\frac{9}{4r}\right)\nonumber\\
& & +a_{2}^{2}\left(\frac{9}{4r}\right)+a_{1}^{4}\left(\frac{81}{32r^{3}}\right)-a_{1}^{2}a_{2}\left(\frac{27}{8r^{2}}\right).
\label{eq:entropyExpandCan}
\end{eqnarray}

Similarly in the expansion of the energy, $\epsilon$, we consider
the interaction energy of A and B particles given by the first term
in the RHS of Eq. (\ref{eq:enenrgyTerm}):
\begin{equation}
\epsilon_{AB}=\frac{r^{2}}{18}-\frac{\sqrt{3}}{8\pi}a_{1}^{2}+\frac{\sqrt{3}}{16\pi}a_{2}^{2}.
\end{equation}
From symmetry, $\epsilon_{AB}=\epsilon_{BC}=\epsilon_{CA}$. Adding the constant term
$-r^{2}/6$, we find:
\begin{eqnarray}
\epsilon & = & -a_{1}^{2}\left(\frac{3\sqrt{3}}{8\pi}\right)+a_{2}^{2}\left(\frac{3\sqrt{3}}{16\pi}\right).
\label{eq:energyExpandCan}
\end{eqnarray}
The critical expansion of the free energy in Eq. (\ref{eq:Fexpand}) is then obtained by
inserting Eqs. (\ref{eq:entropyExpandCan}) and (\ref{eq:energyExpandCan}) into
$\mathcal{F}= \beta \epsilon-s$.

\section{Critical expansion of the nonconserving free energy}

\label{sec:ExpandG}
We present the critical expansion of the free energy of the nonconserving model
whose Hamiltonian  (\ref{eq:gammaH}) is given by
\begin{eqnarray}
\mathcal{H}_{GC,\gamma}\left(\left\{ X_{i}\right\} \right)
= \mathcal{H}\left(\left\{ X_{i}\right\} \right)
 -\gamma N\left(N-1\right)-\mu NL.
\end{eqnarray}
This is a modified Hamiltonian for the case of nonlocal dynamics (see Secition \ref{sec:NonlocalDynamics}).
It reduces to the Hamiltonian (\ref{eq:H_ABCgen}) considered in
Section \ref{sec:PDnonConserving} by setting $\gamma =\frac{1}{6}$.
The free energy of the model is thus

\begin{equation}
\mathcal{G}_{\gamma}\left[\rho_{n}(x)\right]= \beta
\left[\epsilon\left[\rho_{n}(x)\right] + \left(\frac{1}{6}-\gamma\right)r^2-\mu r\right]
- s\left[\rho_{n}(x)\right],
\end{equation}
where entropy and energy,  $s$ and $\epsilon$, are given in Eqs.
(\ref{eq:entropyTerm}) and (\ref{eq:enenrgyTerm}) respectively. We
follow the same expansion procedure presented in \ref{sec:ExpandF},
where the free energy is written as a power series of $a_1$. The
analysis of the multi-critical point of the nonconserving phase
diagram requires terms up to order $a_1^8$. We carried out the
calculation to this order. However, in order to avoid lengthy
expressions, we present it here only up to order $a_1^6$, where the
free energy is expanded in terms that involve only the amplitudes $a_1$ and $a_2$.
 Here we also take into account fluctuations in the total particle density, $r$,
denoted by $\delta r$. The density profile is thus expressed by the
Fourier expansion,
\begin{eqnarray}
\rho_{A}(x)=\frac{r}{3}+\frac{\delta r}{3}+a_{1}\cos\left(2\pi x\right)+a_{2}\cos\left(4\pi x\right).
\end{eqnarray}

The calculation is similar to that detailed in \ref{sec:ExpandF},
with some modifications. First, the  dependence of the entropy of
the vacancies on the density, and thus on $\delta r$, has to be
taken in to account. It takes the form:
\begin{eqnarray}
-s_{0} & = & \left(1-r\right)\ln\left(1-r\right) - \delta r\left[1+\ln\left(1-r\right)\right] \nonumber \\
& +& \frac{\left(\delta r\right)^{2}}{2\left(1-r\right)}+\frac{\left(\delta r\right)^{3}}{6\left(1-r\right)^{2}}.
\end{eqnarray}
In addition,  the density in the homogeneous steady state depends on
the value of $\mu$ and is determined by the equilibrium condition:
\begin{equation}
\frac{\partial}{\partial r}\mathcal{G}_{\gamma,{\rm h}}\left(r\right)=\ln\left(\frac{r}{3\left(1-r\right)}\right)+2\beta\left(\frac{1}{6}-\gamma\right)r-\beta\mu=0,
\end{equation}
where $\mathcal{G}_{\gamma,{\rm h}}$ is the free energy of the homogeneous profile (\ref{eq:Gh_gamma}).
The expansion of the free energy is:
\begin{eqnarray}
\label{eq:Ggamma-expand}
\mathcal{G}_\gamma\left[\rho_{n}(x)\right] & = & \mathcal{G}_{\gamma,{\rm h}} \left(r\right)+\left(\frac{9}{4r}-\frac{3\sqrt{3}\beta}{8\pi}\right)a_{1}^{2}
 +\left(\frac{9}{4r}+\frac{3\sqrt{3}\beta}{16\pi}\right)a_{2}^{2} \nonumber \\
& - & \frac{9}{4r^{2}}a_{1}^{2}\delta r
  + \left(\frac{\beta}{6}-\beta\gamma+\frac{1}{2\left(1-r\right)}+\frac{1}{2r} \right)
\left(\delta r\right)^{2}-\frac{27}{8r^{2}}a_{1}^{2}a_{2} \nonumber \\
& + & \frac{81}{32r^{3}}a_{1}^{4}+\frac{243}{32r^{5}}a_{1}^{6}
-\frac{9}{4r^{2}}a_{2}^{2}\delta r +\frac{81}{8r^{3}}a_{1}^{2}a_{2}^{2} \\
& + & \frac{9}{4r^{3}}a_{1}^{2}\delta r^{2}-\frac{243}{32r^{4}}a_{1}^{4}\delta r
-\frac{243}{16r^{4}}a_{1}^{4}a_{2} \nonumber \\
& + & \frac{27}{4r^{3}}a_{1}^{2}a_{2}\delta r +\left(\frac{1}{6\left(1-r\right)^{2}}-
\frac{1}{6r^{2}}\right)\left(\delta r\right)^{3}
+\mathcal{O}\left(a_{1}^{8}\right) \nonumber ,
\end{eqnarray}

We use the following expansion for the amplitudes:
\begin{eqnarray}
\delta r = A_{0,2}a^{2}_1+A_{0,4}a^{4}_1 , \quad a_{2} = A_{2,2}a^{2}_1+A_{2,4}a^{4}_1.
\label{eq:coeffsCDE}
\end{eqnarray}
Substituting these terms in $\mathcal{G}$ results in the power
series:
\begin{equation}
\mathcal{G}_{\gamma}\left[\rho_{n}(x)\right]=\mathcal{G}_{\gamma,h}\left(r\right)
+g_{2}^{\gamma}a^{2}_1+g_{4}^{\gamma}a^{4}_1+g_{6}^{\gamma}a^{6}_1+\mathcal{O}\left(a^{8}_1\right),
\end{equation}
where
\begin{equation}
g_{2}^{\gamma}=\frac{9}{4r}-\frac{3\sqrt{3}\beta}{8\pi},
\end{equation}
and
\begin{eqnarray}
g_{4}^{\gamma}=\frac{81}{32r^{3}}-\frac{27}{8r^{2}}A_{2,2}-\frac{9}{4r^{2}}A_{0,2}\nonumber\\
+\left(\frac{3\sqrt{3}\beta}{16\pi}+\frac{9}{4r}\right)A_{2,2}^{2}+\left(\frac{\beta}{6}-\beta\gamma
+\frac{1}{2r(1-r)}\right)A_{0,2}^{2}.
\end{eqnarray}
The coefficient $g_{6}^{\gamma}$ can be expressed in a similar fashion.

The coefficients $\left\{A_{i,j}\right\}$ are derived
from the equilibrium condition:
\begin{equation}
\frac{\partial\mathcal{G}_{\gamma}\left[\rho_{n}(x)\right]}{\partial a_{2}}  =  0\, , \qquad
\frac{\partial\mathcal{G}_{\gamma}\left[\rho_{n}(x)\right]}{\partial\left(\delta r\right)}  =  0.
\end{equation}
Expanding the equation for $a_2$ in powers of $a_1$ and using Eq. (\ref{eq:coeffsCDE}),
we find
\begin{eqnarray}
0=\left(\frac{3\sqrt{3}\beta}{8\pi}A_{2,2}-\frac{27}{8r^{2}}+\frac{9}{2r}A_{2,2}\right)a^{2}_{1}\nonumber\\
+\frac{3}{16\pi r^{4}}\left(-81\pi+108A_{2,2}\pi r+36A_{0,2}\pi r-24A_{2,2}A_{0,2}\pi r^{2}\right.\nonumber\\
\left.+24A_{2,4}\pi r^{3}+2\sqrt{3}\beta A_{2,4}r^{4}\right)a^{4}_{1}+\ldots\label{eq:expanddgda2}
\end{eqnarray}
Each power of $a_1$ has to be equal to zero independently. From the second
order term, we find
\begin{equation}
A_{2,2}=\frac{9\pi}{r\left(12\pi+\sqrt{3}\beta r\right)}.
\label{A22}
\end{equation}
Similarly, expanding the equation
$\partial\mathcal{G}_{\gamma}/\partial\left(\delta r\right)=0$ in
powers of $a_1$ one finds
\begin{equation}
A_{0,2}=\frac{27\left(1-r\right)}{4r\left[3+\beta
r\left(1-r\right)\left(1-6\gamma\right)\right]}. \label{A02}
\end{equation}
Equations (\ref{A22}) and (\ref{A02}) are then used to evaluate
$A_{2,4}$ from the fourth order term in (\ref{eq:expanddgda2}):
\begin{eqnarray}
A_{2,4}=\frac{81\pi}{2r^{3}\left(12\pi+\sqrt{3}\beta r\right)^{2}}\times\nonumber\\
\frac{\sqrt{3}\beta r^{2}\left[3+\beta(1-r)(1-6\gamma)\right]-18\pi (1-r)}{3+\beta r\left(1-r\right)\left(1-6\gamma\right)}.
\end{eqnarray}
The higher order coefficients are found in a similar manner.

Substituting the coefficients in the expression for the free energy
$\mathcal{G}_{\gamma}$, we obtain
\begin{equation}
g_{4}^{\gamma}=\frac{81}{32r^{3}}\left[\frac{\sqrt{3}\beta r+6\pi}{\sqrt{3}\beta r+12\pi}-\frac{3\left(1-r\right)}{3+\beta\rho\left(1-r\right)\left(1-6\gamma\right)}\right].
\end{equation}
To avoid lengthy expressions we display the expression for $g_{6}$
only along the critical line $\beta_{c}=2\pi\sqrt{3}/r$. With the
notation $\theta=\left(1-r\right)\left(1-6\gamma\right)$ it is given
by:
\begin{eqnarray}
g_{6}^{\gamma}\left(\beta_{c}\right)=\frac{243}{64r^{5}\left[3+2\pi\sqrt{3}\theta\right]^{3}}\times\nonumber\\
\left[16\pi^{3}\sqrt{3}\theta^{3}+6\pi^{2}\theta^{2}\left(17r-5\right)\right.\nonumber\\
\left.+6\pi\sqrt{3}\theta\left(9r^{2}-r-2\right)+9\left(6r^{2}-5r+1\right)\right].
\end{eqnarray}
For $\gamma=\frac{1}{6}$ we find that at the multicritical point,
where $g_{2}^{\gamma}=g_{4}^{\gamma}=0$, one also has
$g_{6}^{\gamma}=0$. We therefore need to calculate the eighth-order
coefficient, $g_{8}^{\gamma}$, in the same manner described above.
This requires the evaluation of the amplitude $a_4$ as well. This
calculation, whose details are not presented here, yields the
following expression for $g_{8}^{\gamma}$ along the critical line
\begin{eqnarray}
 g_{8}^{\gamma}\left(\beta_{c}\right)=\frac{243}{1024r^{7}\left[3+2\pi\sqrt{3}\theta\right]^{5}}\times\nonumber\\
\left[5632\pi^{5}\sqrt{3}\theta^{5}+24\pi^{4}\theta^{4}\left(2883r-1123\right)\right.\nonumber\\
+48\pi^{3}\sqrt{3}\theta^{3}\left(1800r^{2}-717r-203\right)\nonumber\\
+72\pi^{2}\theta^{2}\left(1458r^{3}+567r^{2}-1413r+268\right)\nonumber\\
+18\pi\sqrt{3}\theta\left(3645r^{3}-3186r^{2}+204r+217\right)\nonumber\\
\left.+27\left(1215r^{3}-1692r^{2}+762r-109\right)\right].
\end{eqnarray}
The coefficient $g_{4}^{\gamma},g_{6}^{\gamma}$ and $g_{8}^{\gamma}$
are used for the expansion of the free energy for $\gamma=\frac{1}{6}$ in Eq. (\ref{eq:GExpand})
and for  $\gamma\neq\frac{1}{6}$ in Eq. (\ref{eq:GExpand_gamma}).

\section{Steady-state profiles in the continuum limit}
\label{sec:AnalyticSolution}

In order to locate the first order transition line of the ABC model one
has to calculate the density profiles of the three species in the
ordered phase. In this Appendix we provide an analytic solution for
the profiles by applying the approach introduced in \cite{Ayyer2009}.
This is done first by translating the dynamical rules in Eq.
(\ref{eq:CanonicalDynamics}) to an equation for the time evolution
of $\left\langle A_{i}\right\rangle$ as
\begin{eqnarray}
\frac{d }{dt}\left\langle A_{i}\right\rangle=q\left\langle
A_{i-1}B_{i}\right\rangle +q\left\langle C_{i}A_{i+1}\right\rangle
+\left\langle B_{i}A_{i+1}\right\rangle +\left\langle
A_{i-1}C_{i}\right\rangle \nonumber \\-q\left\langle
A_{i}B_{i+1}\right\rangle -q\left\langle C_{i-1}A_{i}\right\rangle
-\left\langle B_{i-1}A_{i}\right\rangle -\left\langle
A_{i}C_{i+1}\right\rangle. \label{eq:CorrelatorsDynamics}
\end{eqnarray}
The corresponding equation for $\left\langle
B_{i}\right\rangle,\left\langle C_{i}\right\rangle$ are obtained by
cyclic permutation of $A,\,B$ and $C$. As has been shown in
\cite{Ayyer2009}, in the weak asymmetry limit and for large $L$ one has
\begin{equation}
 \langle X_iZ_{i\pm 1}\rangle = \langle X_i \rangle \langle Z_{i\pm 1}
\rangle + \mathcal{O}(\frac{1}{L}), \label{eq:meanfield}
\end{equation}
 where $X$ and $Z$ are either $A,\,B$ or $C$.
In the continuum limit \cite{Clincy2003,Fayolle2004,Ayyer2009} one can write
\begin{equation}
\left\langle A_{i\pm1}\right\rangle=
\rho_{A}\pm\frac{1}{L}\frac{\partial\rho_A}{\partial x} + \mathcal{O}(\frac{1}{L^2})
\label{eq:derivative}
\end{equation}
and similarly for $B$ and $C$.
Using (\ref{eq:meanfield}) and (\ref{eq:derivative})
 while keeping only leading terms,
 Eq. (\ref{eq:CorrelatorsDynamics}) becomes:
\begin{eqnarray}
\frac{\partial\rho_{A}}{\partial\tau}= \beta \frac{\partial}{\partial x} \left[\rho_{A}\left(\rho_{B}-\rho_{C}\right)\right]+\frac{\partial^{2}\rho_{A}}{\partial x^{2}},
\label{eq:MeanField}
\end{eqnarray}
where $\tau=t/L^2$ and $q$ has been replace by $e^{-\beta/L}\simeq 1-\frac{\beta}{L}$.
The first term of RHS of Eq. (\ref{eq:MeanField}) accounts for the drive which favors an ordered phase,
whereas the second term represents the diffusion which favors a homogeneous phase.
 In the weak asymmetry limit, the two terms are comparable in magnitude
and thus compete one another.

These hydrodynamic equations are in fact coupled Burgers equations, whose stationary solution obeys
\begin{eqnarray}
\label{eq:IntegratedMF}
\frac{\partial\rho_{A}}{\partial x}=-\beta \left[\rho_{A}\left(\rho_{B}-\rho_{C}\right)\right] \nonumber \\
\frac{\partial\rho_{B}}{\partial x}=-\beta \left[\rho_{B}\left(\rho_{C}-\rho_{A}\right)\right] \\
\frac{\partial\rho_{C}}{\partial x}=-\beta \left[\rho_{C}\left(\rho_{A}-\rho_{B}\right)\right], \nonumber
\end{eqnarray}
obtained by setting the LHS of Eq. (\ref{eq:MeanField}) to zero and integrating over $x$.
The absence of integration constant is due to the fact that there are no steady-state
currents in the case of equal densities.

Eqs. (\ref{eq:IntegratedMF}) have been solved by Ayyer et al. \cite{Ayyer2009} for the ABC model on
an interval. For equal densities their derivation applies for
 periodic boundary condition as well. Multiplying the three equations by $\rho_B\rho_C$, $\rho_A\rho_C$
and $\rho_A\rho_B$, respectively, and summing the resulting equations, yields
$\frac{d}{dx}(\rho_A\rho_B\rho_C)=0$, and consequently
\begin{equation}
\rho_A(x)\rho_B(x)\rho_C(x)=K,
\label{eq:KDef}
\end{equation}
where $K>0$ is a constant.
Equation (\ref{eq:KDef}) in conjunction with $\rho_A+\rho_B+\rho_C=1$ decouples Eqs.
(\ref{eq:IntegratedMF}), yielding an equation for $\rho_A$ :
\begin{equation}
\label{eq:SignleFieldMF}
\frac{\partial\rho_{A}}{\partial x}=\pm\beta \sqrt{\rho_{A}^{2}(1-\rho_{A})^{2}-4K\rho_{A}}.
\end{equation}
and similarly for  $\rho_B$ and $\rho_C$. In terms of
the rescaled variables $t=2\beta x $ and $y(t)=\rho_A(x)$
Eq. (\ref{eq:SignleFieldMF}) may be written as
\begin{equation}
\label{eq:particleEOM}
\frac{1}{2}\left[\frac{d y(t)}{dt}\right]^2+U_K\left[ y(t) \right]=0,
\end{equation}
where
\begin{equation}
U_K(y)=\frac{1}{2}K y-\frac{1}{8}y^{2}(1-y)^{2}.
\end{equation}
This equation can be viewed as an
equation of motion of a zero-energy particle with mass $1$ in a quartic potential.
 The four roots of $U_K(y)=0$ are denoted here as
 $\left\{0,a,b,c\right\}$. They are functions of $K$ obeying $0<a<b<1<c$.
 The physical trajectory, where $0\leq y(t) \leq 1$ and $U_K(y) < 0$, is the one where
 the particle oscillates between $a$ and $b$ with a period of
\begin{equation}
T=2\intop_{a}^{b}\frac{dy}{\sqrt{-2U_{K}(y)}}.
\label{eq:ParticlePeriod}
\end{equation}

Based on the free energy of the ABC model, it can be shown that Eq. (\ref{eq:particleEOM})
has a unique steady-state solution given by $T=2\beta$, and
many quasi-stationary solutions given by $T=2\beta m$ where $m>1$ is an integer. The trajectories
are unique up to the choice of initial time for the motion of the particle which corresponds to the translation symmetry of the
profile.
In order to obtain an analytic expression for the integral in Eq. (\ref{eq:ParticlePeriod})
it is convenient to rewrite it as an elliptic integral of the first kind of the form
\begin{equation}
F\left(x,k\right)=\intop_{0}^{x}\frac{dz}{\sqrt{\left(1-z^2\right)
\left(1-k^2z^2\right)}}.
\label{eq:elliptic_F}
\end{equation}
This is done using a M{\"o}bius transformation that
takes the roots of the potential, $U_K(y)$, from $\left\{0,a,b,c\right\}$
 to $\left\{-1,-1/k,1/k,1\right\}$, which are the roots of
 the denominator of (\ref{eq:elliptic_F}). The transformation is given by
\begin{eqnarray}
z=f\left(y\right)=\frac{\alpha_+y-1}{\alpha_-y+1},
\end{eqnarray}
where
\begin{equation}
\alpha_{\pm}=\frac{\pm ab+\sqrt{ab\left(c-b\right)\left(c-a\right)}}{abc}.
\end{equation}
and
\begin{equation}
k=\frac{1+\alpha_{-}a}{1-\alpha_{+}a}.
\end{equation}
The parameters $\alpha_{-}$,$\alpha_{+}$ and $k$ are functions of $K$ through $a,b$ and $c$.
Let $t(y)$ be the time it takes the particle to move from $a$ to $y$. Using the transformation above
it may be expressed as
\begin{eqnarray}
t &= &2\intop_{a}^{y}\frac{dy'}{\sqrt{-2U_{K}(y')}} = \nonumber \\
& = & \varkappa \intop^{f(y)}_{-1/k}\frac{dz}{\sqrt{(1-z^2)(1-k^2z^2)}}=
\varkappa \left[ F\left(1/k,k\right)+F\left(f(y),k\right) \right],
\label{eq:HyberbplicTime}
\end{eqnarray}
where
\begin{equation}
\varkappa = \frac{2(\alpha_{+}+\alpha_{-})}
{\sqrt{(1-\alpha_{+}a)(1-\alpha_{+}b)(1-\alpha_{+}c)}}.
\end{equation}
The full period is given by setting $f(y)=1/k$.
The condition of $T=2\beta$ yields an equation that connects
$\beta$ and $K$ through $k(K)$ and $\varkappa(K)$:
\begin{equation}
\beta=2\varkappa F\left(1/k,k\right).
\end{equation}

With $K(\beta)$ known, we can proceed to express the profile by inverting
Eq. (\ref{eq:HyberbplicTime}). This is done using the Jacobi elliptic function, $\mathrm{sn}\left(x,k\right)$,
defined by the equation $F\left(\mathrm{sn}\left(x,k\right),k\right)=x$. The resulting profile is given up to
translations of $x$ as
\begin{equation}
\rho_A(x)=\frac{1+\mathrm{sn}\left(2\beta x / \varkappa,k \right) }
{\alpha_+ - \alpha_- \mathrm{sn}\left(2\beta x / \varkappa,k \right)}.
\end{equation}
The profiles for the two other species are $\rho_B(x)=\rho_A(x-\frac{1}{3})$ and
$\rho_C(x)=\rho_A(x+\frac{1}{3})$.

The solution above of the standard ABC model with equal densities can easily be
extended to the conserving model which includes vacancies.
 This is done by applying the mapping between the steady
 state of an $L$-size generalized ABC model to a 'condensed' $N$-size system without
 vacancies but with an inverse temperature of $\beta r$
 (see the first paragraph of Section \ref{sec:PDconserving}).
The mapping back to the $L$-size system requires the addition
 of vacancies into the lattice, done by multiplying the profile by $r$.
 Hence, the steady-state profile of the conserving model is given as
\begin{equation}
\rho_A(x)=r\frac{1+\mathrm{sn}\left(2\beta r x / \varkappa,k \right) }
{\alpha_+ - \alpha_- \mathrm{sn}\left(2\beta r x / \varkappa,k \right)},
\label{eq:analytic_profile_app}
\end{equation}
where $\varkappa\,,k\,,\alpha_{+}$ and $\alpha_{-}$ are now functions of $K(\beta r)$, and the two other
profiles are again given as $\rho_B(x)=\rho_A(x-\frac{1}{3})$ and
$\rho_C(x)=\rho_A(x+\frac{1}{3})$.

We now consider the steady-state profile of the nonconserving model.
The corresponding mean-field are obtained in a similar way by
 including the nonconserving process in Eq.  (\ref{eq:addremoverates}), yielding
\begin{equation}
\frac{\partial\rho_{A}}{\partial\tau}= I_1+pL^2I_2,\label{eq:MeanField1}
\end{equation}
where $I_1= \beta \frac{\partial}{\partial x} \left[\rho_{A}\left(\rho_{B}-\rho_{C}\right)\right]+\frac{\partial^{2}\rho_{A}}{\partial x^{2}}$
 represents the drive and diffusion, and $I_2=\rho_0^3-e^{-3\beta\mu}\rho_A\rho_B\rho_C$ is
 corresponds to the evaporation and
deposition processes. The equations for $B$ and $C$ are again given by cyclic permutations of
 this equation over $A,\,B$ and $C$. Detailed balance with respect to the conserving
 (\ref{eq:CanonicalDynamics}) and nonconserving (\ref{eq:addremoverates}) processes
implies that in the steady state $I_1=0$ and $I_2=0$ independently.
 Setting $I_1=0$  yields the same
stationary equation and thus the same profile as in the conserving model. Since
this profile obeys $\rho_A(x)\rho_B(x)\rho_C(x)=Kr^3$ and $\rho_0(x)=1-r$, $I_2$ can
 indeed be set to zero for all $x$, yielding a relation between $r$ and $\mu$ :
\begin{equation}
\mu=\frac{1}{\beta}\ln{\left[\frac{rK^{1/3}(\beta r)}{1-r}\right]}.
\label{eq:analytic_mu_app}
\end{equation}
This seemingly accidental coincidence is due to the choice of a nonconserving process that
 maintains detailed balance.

Because both the conserving and nonconserving models
have the same stationary profiles, we can use Eq. (\ref{eq:analytic_mu_app})
to compute the chemical potential in the conserving model. For certain values of
$r$ we find that this definition yields a negative compressibility,
 $\frac{\partial\mu}{\partial r}<0$. This profile is thus
unstable in the nonconserving model, giving rise to inequivalence of ensembles.

\bibliographystyle{iopart-num}
\bibliography{ABCModel}

\end{document}